\documentclass[acmsmall]{acmart}
\usepackage[utf8]{inputenc}
\usepackage[english]{babel}
\usepackage{glossaries}
\usepackage{enumitem}
\usepackage{graphicx}
\usepackage{csquotes}
\usepackage{url}
\usepackage{fmtcount}
\usepackage{pbox}
\usepackage{listings}
\usepackage{eiffel}
\usepackage{boogie}
\usepackage{array,graphicx}
\usepackage{booktabs}
\usepackage{pifont}
\usepackage{hhline}
\usepackage{adjustbox}
\usepackage{multirow}
\usepackage{spverbatim}
\usepackage{alloy-style}
\usepackage{vdmlisting}
\usepackage{adjustbox}
\usepackage{amsmath, amssymb}
\usepackage{hyperref}
\usepackage{caption}
\newcommand*\rot{\rotatebox{70}}
\newcommand*\OK{\ding{51}}
\newcommand*\NOK{\ding{56}}

\usepackage{color}      % for comments
\usepackage{amssymb}    % for comments
\newcommand{\mynote}[3][black]{\textcolor{#1}{\fbox{\bfseries\sffamily\scriptsize{#2}}
{\small$\blacktriangleright$\textsf{\emph{#3}}$\blacktriangleleft$}}}

\newcommand{\TODO}[1]{\mynote[red]{TODO}{#1}}

\renewcommand{\TODO}[1]{} % to remove all the ToDos

\newcommand{\mysec}[1]{section~\ref{sec:#1}}
\newcommand{\myfig}[1]{Fig.~\ref{fig:#1}}
\newcommand{\mytab}[1]{Table~\ref{tab:#1}}

\newcommand{\e}[1]{\mbox{\lstinline[basicstyle=\normalsize]|#1|}}

\title{The role of formalism in system requirements (\textit{full~version})}
\author{Jean-Michel Bruel}
\email{bruel@irit.fr}
\affiliation{%
  \institution{University of Toulouse, IRIT}
}

\author{Sophie Ebersold}
\email{sophie.ebersold@irit.fr}
\affiliation{%
  \institution{University of Toulouse, IRIT}
}

\author{Florian Galinier}
\email{florian.galinier@irit.fr}
\affiliation{%
  \institution{University of Toulouse, IRIT}
}

\author{Manuel Mazzara}
\email{m.mazzara@innopolis.ru}
\affiliation{%
    \institution{Innopolis University}
}

\author{Alexandr Naumchev}
\email{a.naumchev@innopolis.ru}
\affiliation{%
    \institution{Innopolis University}
}

\author{Bertrand Meyer}
\email{Bertrand.Meyer@inf.ethz.ch}
\affiliation{%
    \institution{Schaffhausen Institute of Technology and IRIT}
}
\begin{document}

\renewcommand*{\CustomAcronymFields}{%
  name={\the\glsshorttok},% name is abbreviated form
  description={\the\glslongtok},% description is long form
  first={\noexpand\the\glsshorttok\space(\the\glslongtok)},
  firstplural={\noexpand\emph{\the\glslongtok\noexpand\acrpluralsuffix}\space(\the\glsshorttok)},%
  text={\the\glsshorttok},%
  plural={\the\glsshorttok\noexpand\acrpluralsuffix}%
}
\SetCustomStyle
% It's a good idea to add them by alphabetic order to quickly see that one is missing...

\newacronym{ACRE}{ACRE}{Approach for Context-based Requirement Engineering}
\newacronym{BON}{BON}{Business Object Notation}
\newacronym{CAS}{CAS}{Complex Adaptive System}
\newacronym{CSP}{CSP}{Constraint Satisfaction Problem}
\newacronym{DBC}{DbC}{Design by Contract}
\newacronym{DSL}{DSL}{Domain Specific Language}
\newacronym{DSML}{DSML}{Domain Specific Modeling Language}
\newacronym{EAST-ADL2}{EAST-ADL2}{Electronic Architecture \& Software Tools - Architecture Description Language}
\newacronym{GEMOC}{GEMOC}{Generic Model of Computation}
\newacronym{GORE}{GORE}{Goal-Oriented Requirements Engineering}
\newacronym{HDL}{HDL}{Hardware Description Language}
\newacronym{INCOSE}{INCOSE}{International Council on Systems Engineering}
\newacronym{LGS}{LGS}{Landing Gear System}
\newacronym{MARTE}{MARTE}{Modeling and Analysis of Real Time and Embedded systems}
\newacronym{MBSE}{MBSE}{Model-Based System Engineering}
\newacronym{MDE}{MDE}{Model Driven Engineering}
\newacronym{NL}{NL}{Natural Language}
\newacronym{OCL}{OCL}{Object Constraint Language}
\newacronym{RE}{RE}{Requirements Engineering}
\newacronym{SysML}{SysML}{Systems Modeling Language}
\newacronym{SoS}{SoS}{System of Systems}
\newacronym{STD}{STD}{State Transition Diagrams}
\newacronym{UML}{UML}{Unified Modeling Language}
\newacronym{URML}{URML}{User Requirements Modeling Language}
\newacronym{VV}{V\&V}{Verification and Validation}

%------------------- abstract -----------------------
\begin{abstract}

% Table approaches / who is in charge

% Context (why do we need this survey?)

A major determinant of the quality of software systems is the quality of their requirements, which should be both understandable and precise. Most requirements are written in natural language, good for understandability but lacking in precision. 

To make requirements precise, researchers have for years advocated the use of mathematics-based notations and methods, known as ``formal''. 
Many exist, differing in their style, scope and applicability. The present survey discusses some of the main formal approaches and compares them to informal methods.

The analysis uses a set of 9 complementary criteria, such as level of abstraction, tool availability, traceability support.  It classifies the approaches into five categories: general-purpose, natural-language, graph/automata, other mathematical notations, seamless (programming-language-based). It presents approaches in all of these categories, altogether 22 different ones, including for example SysML, Relax, Eiffel, Event-B, Alloy.

The review discusses a number of  open questions, including seamlessness, the role of tools and education, and how to make industrial applications benefit more from the contributions of formal approaches. 

\textit{This is the full version of the survey, including some sections and two appendices which, because of length restrictions, do not appear in the submitted version.
}

%-------------------------------------------------------------------------------------

% Definition and understanding of software requirements is fundamental to ensure software quality. Natural language is very expressive but, at the same time, ambiguous and can lead to misunderstanding. Formal techniques promise to overcome this problem at the cost of reducing expressiveness.
% Target and goal
%We conducted a comprehensive survey on requirements techniques, from the more natural language based ones to the more formal ones.

%Definitions (spec vs reqs).
%We distinguish between specification and requirements: a specification describes the technical properties of a system while requirements describe the properties of a system as relevant to its human users.
% Process we've taken to conduct the survey.
%We characterized the surveyed techniques in five distinct categories (general purpose, natural language-based, graph and automata, other mathematical notations and programming language-based) and identified properties that relevant for requirements engineering techniques. 
% Number/kind of approaches surveyed. => 22 representative approaches
%We have examined more than twenty approaches against the identified properties. The an
% Main results and remaining open challenges from the survey.

\end{abstract}
%----------------------------------------------------
\acmMonth{11}
\acmYear{2019}
\maketitle

\section{Introduction}
%----------------------------------------------------

%no question in the entire field of information technology has more import than software quality. 

In a world where software pervades every aspect of our lives, a core issue for the IT industry is how to guarantee the quality of the systems it produces. Software quality is a complex and widely studied topic, but it is not hard to provide a simple definition: quality means that \textit{the software does the right things, and does them right}. These ``things'' that a software system does are known as its \textbf{requirements}. Not surprisingly, requirements engineering is a core area of software engineering.

Both goals, doing the right things and doing things right, fundamentally depend on the quality of the requirements: the requirements must define the system so that it will satisfy user needs; and they must make it possible to assess a candidate implementation against this definition, a task known as \emph{validation} (as distinct from \emph{verification}, which assesses the internal properties of the implementation).

Validation can only be effective if the requirements are precise. Precision in science and technology is typically achieved by using  mathematical methods and notations, also known in software engineering as \emph{formal} methods and notations. This survey examines the state of the art in applying such formal approaches to software requirements, and perspectives for their further development.

Precision is so important that an outsider to the field might assume that \emph{all} software development starts with a formal description of the requirements. This approach is by far not the standard today; most practical projects, if they use explicit requirements at all, describe them informally, either in the form of a natural-language ``requirements document'' or (in agile methods) through individual ``user stories'', also expressed in natural language.

A number of \emph{formal requirements methods} also exist. The following sections present some of them together with the supporting notations and tools, and discuss their applicability as a replacement for, or complement to, the dominant informal approaches.

A version of this article has been submitted for publication, omitting a few sections and the appendices because of length restrictions. To maintain consistency of numbering, the supplementary sections have special numbers such as 4.1.A in the present full-length version.
% %----------------------------------------------------
% \section{Scope}
% %----------------------------------------------------

\subsection{Terminology: requirements versus specifications}

``Formal \emph{specification}'' is a well-accepted concept, covered by various survey articles \cite{tilley2005survey, woodcock2009formal}. The present article surveys formal \textit{requirements}. While it is generally accepted that the concepts of specification and requirements are both distinct and related, there exist -- even within a single normative source such as the Software Engineering Body of Knowledge \cite{IEEEComputerSociety:2014:GSE:2616205} --  varying definitions of the  difference. This article will rely on the following definition of the distinction, with which we believe many knowledgeable professionals would agree.

First, what is common to requirements and specifications: both describe the ``what'' of a system or system element — its purpose and the constraints to which it is submitted — rather than the ``how'', which is the responsibility of other software engineering tasks, \emph{design} and \emph{implementation}.

As to the difference, it is one of purpose and scope:
\begin{itemize}
\item A specification describes the \emph{technical properties} of a system, or often some part of a system. For example, the rendering engine of a Web browser must display any given HTML text in a certain way. A specification will describe the desired properties of this rendering.
\item Requirements describe the properties of a system or system element as relevant to its human users, or more generally its \emph{environment}. (This generalization is necessary since not all systems have a direct human ``user'': many, such as the engine of a car, serve the purpose of another system, in this case the car as a whole.) For example, the requirements for a Web browser describe the functionality that it provides to its users and the constraints on this functionality.
\end{itemize}

This definition indicates why the distinction between requirements and specifications cannot be absolute: to produce the requirements of a complex system, it will be necessary to decompose them into sub-requirements of its components, and the further you go into this decomposition the more detailed and technical the requirements will become, getting closer to specifications. Even if not absolute, the distinction is useful. This article focuses on formal approaches to requirements, applicable to entire systems and covering the properties directly relevant to the its users and environment. While there have been systematic studies of formal specifications, formal requirements are a less well-explored topic. The present article is an attempt to fill this gap. 

\subsection{Terminology: ``specification''} 

Two further observations on terminology will help avoid confusion:
\begin{itemize}
\item The verb ``to specify'' is in common English use to mean ``to describe precisely'' or ``to include a mention of''. Such use is applicable to requirements too, as in ``Your implementation ignores this case, but the requirements document specifies that it must be reported as an error!''. It would be cumbersome to deprive ourselves from such standard usage, as long as it does not imply any confusion with the term ``specification'' in its technical sense discussed above.
\item Following the standard practice of the requirements literature, this article uses ``a \emph{requirement}'' as the description of a particular property of a system and ``the \emph{requirements}'' as the collection of every such individual ``requirement'' for a system. ``Requirements'' here is not just the plural of ``requirement'' but a concept on its own, often understood as an abbreviation for ``the requirements document''. In this discussion, ``requirements'' denotes that collective concept. When the emphasis is on one or more specific ``requirement'' the phrasing will reflect it clearly, as in ``requirement \#25'' or ``the following three requirement \emph{elements}'' etc.
\end{itemize}

\section{Running example}\label{sec:casestudy}
%----------------------------------------------------

To illustrate and compare the approaches surveyed, it is useful to rely on a common example. 
%Need to decide whether we use he Library example or not, because if we do the following sentence needs to be removed/adapted.
A well-known survey that used such an example is Wing's 1988 study of specifications of a simple library system \cite{Wing:1988:SSL:624570.624745}. To reflect the challenges of today's demanding IT applications, we need a more difficult example. This article uses the Landing Gear System for airplanes (LGS), a case study \cite{boniol_landing_2014} that has received wide attention in the requirements literature (including in the authors' own previous work \cite{NAUMCHEV2019131}). The LGS is a a complex, critical system for which the requirements involve diverse stakeholders and many fields of expertise.

Physically, an LGS consists of the landing set, a gear box that stores the gear in retracted position, and a door attached to the box. 
The door and gear are independently activated by a digital controller. The controller reacts to changes of position of a handle by initiating either gear extension or retraction process. 
In other words, the controller must align in time the events of changing the handle’s position and sending commands to the door and the gear actuators: doors are opened and closed and gears are either moving out (extension) or moving in (retraction).
One may express these rules, still in natural language but more precisely, as follows (all applicable to ``command line'' mode): 	 
\begin{itemize}
\item (R11bis) If the landing gear command handle has been pushed down and stays down, then eventually the gears will be locked down and the doors will be seen closed. 
\item (R12bis) If the landing gear command handle has been pushed up and stays up, then eventually the gears will be locked retracted and the doors will be seen closed.
\item (R21) If the landing gear command handle remains in the down position, then retraction sequence is not observed.
\item (R22) If the landing gear command handle remains in the up position, then outgoing sequence is not observed.
\end{itemize}

%We will see how to express some of these properties (a small subset of the entire LGS description) in the approaches surveyed.

We will see how to express some of these properties in some of the approaches surveyed.

%----------------------------------------------------
\section{Classifying approaches to requirements}\label{sec:classification}
%----------------------------------------------------

This section introduces a classification of  approaches into five categories (\ref{sec:sec:classification_criteria}), introduces the approaches retained and why they were retained (\ref{sec:sec:choice_criteria}), and lists the assessment criteria (\ref{sec:sec:criteria})

\subsection{The classification}
\label{sec:sec:classification_criteria}

Approaches fall into five categories based on how they express requirements:
\begin{itemize}
\item \textit{Natural language} category approaches express requirements in English or another human language, although they can restrict the degree of ``naturalness'' of the text.
\item\textit{Semi-formal} approaches codify the form of requirements, in effect defining a precise requirements language, which is neither a mathematical notation (as in the next two categories) nor derived from an programming language (as in the last category).
\item\textit{Automata, graphs} approaches rely on notations based on automata or graph theory. They usually provide graphical support and may leave the mathematical basis implicit since graphs, in particular, can be used a diagrammatic tool without deep mathematical knowledge.
\item\textit{Mathematical} approaches rely on mathematical formalisms other than those of the previous category.The theoretical basis is, generally, mathematical logic.
\item\textit{Seamless} approaches integrate requirements closely with other software tasks (design, implementation...), using a programming language as notation.
\end{itemize}

\subsection{Selection criteria and list of approaches surveyed}
\label{sec:sec:choice_criteria}

The field of requirements engineering is rich with methods and tools. Any survey must involve a choice and, inevitably, some subjectivity. The present discussion has retained approaches that meet one or more of the following criteria:
\begin{itemize}
\item Widely used (as in the case of commercially available tools such as Doors).
\item Widely publicized.
\item Influential.
\item Possessing, in the authors' view, other distinctive characteristics that warrant discussion.
\end{itemize}
Note that some of the authors have (separately) been involved in three of the methods reviewed, Relax (\ref{sec:sec:sec:Relax}), multirequirements (\ref{sec:sec:sec:multirequirements}), and seamless object-oriented requirements (SOOR) (\ref{sec:sec:sec:seamless_requirements}).

\mytab{approach_list} lists the approaches retained for this survey. The references to the corresponding publications also appear here. When these publications do not give an approach an explicit name, we devised one carrying the central concept, such as Requirements Grammar, and marked it with an asterisk*.

\begin{table}[ht!] 
\centering
    \begin{tabular}{| l | l | l | l |}
    \hline
       Name &  Category & References & Section \\ 
    \hhline{|=|=|=|=|}
       %\cmidrule{2-12}

%--- Natural language
Requirements Grammar* & \multirow{6}{*}{Natural language} &  \cite{scott_context-free_2004} & \ref{sec:sec:sec:RequirementsGrammar}\\ 
    \cline{1-1}\cline{3-4}
Relax & & \cite{whittle_relax:_2009}  & \ref{sec:sec:sec:Relax}\\ 
	\cline{1-1}\cline{3-4}
Stimulus & &  \cite{jeannet_debugging_2015}  & \ref{sec:sec:sec:Stimulus}\\
	\cline{1-1}\cline{3-4}
NL to OCL* &  & \cite{hahnle_authoring_2002} & 
%\ref{sec:sec:sec:NLtoOCL}
\hyperref[sec:sec:sec:NLtoOCL]{4.1.A}\\
    \cline{1-1}\cline{3-4}
NL to STD* &   & \cite{aceituna_evaluating_2011}  &
%\ref{sec:sec:sec:NLtoSTD}
\hyperref[sec:sec:sec:NLtoSTD]{4.1.B}\\
    \cline{1-1}\cline{3-4}
NL to OWL* & & \cite{li_stakeholder_2015}  & \ref{sec:sec:sec:NLtoOWL}\\ 
    \hhline{|=|=|=|=|}
%--- Semi-formal
Doors   & \multirow{6}{*}{Semi-formal}  & \cite{DOORS}  &\ref {sec:sec:sec:DoorsReqtify}\\
    \cline{1-1}\cline{3-4}
Reqtify & & \cite{Reqtify}  &\ref{sec:sec:sec:DoorsReqtify}\\
 \cline{1-1}\cline{3-4}
 KAOS & & \cite{lamsweerde_goal-oriented_2001}  &
 \ref{sec:sec:sec:kaos}\\
    \cline{1-1}\cline{3-4}
URN &  & \cite{Amyot2003}  & 
%\ref{sec:sec:sec:urn}
%4.2.A (online version)\\ 
\hyperref[sec:sec:sec:urn]{4.2.A}\\
    \cline{1-1}\cline{3-4}
SysML &  & \cite{omg_omg_2007}  & \ref{sec:sec:sec:sysml}\\ 
    \cline{1-1}\cline{3-4}
URML &  & \cite{berenbach_use_2012}  & 
%URML &  & \cite{helming_towards_2010, berenbach_use_2012}  & %\ref{sec:sec:sec:urml}
\hyperref[sec:sec:sec:urml]{4.2.B}\\
    \cline{1-1}\cline{3-4}

    \hhline{|=|=|=|=|}
%--- Graph and Automata
Petri Nets &\multirow{5}{*}{Automata- or graph-based} & \cite{petrinets}  & 
%\ref{sec:sec:sec:petrinets}
\hyperref[sec:sec:sec:petrinets]{4.3.A}\\ 
    \cline{1-1}\cline{3-4}
Statecharts &  & \cite{Harel87}  & \ref{sec:sec:sec:statecharts}\\
    \cline{1-1}\cline{3-4}
Problem Frames &  & \cite{Jackson:2000}  & \ref{sec:sec:sec:problem_frames}\\ 
    \cline{1-1}\cline{3-4}
FSP/LTSA &  & \cite{LTSA}  & \ref{sec:sec:sec:ltsa}\\ 
    \cline{1-1}\cline{3-4}
FORM-L &   & \cite{nguyen_verification_2015}  & \ref{sec:sec:sec:form_l}\\
    \hhline{|=|=|=|=|}
%--- Mathematical notation
Event-B & \multirow{5}{*}{Mathematical notation} & \cite{Abrial2010}  & \ref{sec:sec:sec:event_b}\\ 
    \cline{1-1}\cline{3-4}
VDM  &  & \cite{BjornerJones78}  & 
\hyperref[sec:sec:sec:VDM]{4.4.A}\\ 
    \cline{1-1}\cline{3-4}
Process Algebra &  & \cite{Hoare:1978, Milner:1982}  &
%\ref{sec:sec:sec:process_algebra}
\hyperref[sec:sec:sec:process_algebra]{4.4.B}\\ 
    \cline{1-1}\cline{3-4}
Alloy &  &  \cite{Jackson:2006}  & \ref{sec:sec:sec:alloy}\\ 
    \cline{1-1}\cline{3-4}
Tabular Relations &  & \cite{parnas_precise_2011}  &
\hyperref[sec:sec:sec:tabular_relations]{4.4.C}\\ 
    \hhline{|=|=|=|=|}
%--- Programming Language
Multirequirements & Seamless & \cite{meyer_multirequirements_2013, Naumchev2017} & \ref{sec:sec:pl_based}\\ 
    \hline
       % \cmidrule[1pt]{2-12}
\end{tabular}
\caption{Requirements approaches surveyed}
\label{tab:approach_list}
\end{table}

%\mm{TODO: I think we should try to describe the rationale for the categorization}

%Each notation or method will be reviewed according to the following properties/attributes (ordered to make the narrative smooth). ABOUT HALF A PAGE DESCRIPTION FOR EACH NOTATIONS/METHODS. ADD BIBLIOGRAPHY ITEMS FOR EACH OF THEM.

%\begin{enumerate}
%\item System vs environment
%\item Intended audience
%\item Level of abstraction
%\item Associated method
%\item Traceability support
%\item Non-functional requirements
%\item Semantic definition
%\item Tool support
%\item Verifiability
%\end{enumerate}

%\jmb{I thought we'd use those properties instead of those in \mysec{properties}.}

%----------------------------------------------------
\subsection{Criteria for assessing approaches}
\label{sec:sec:criteria}
%----------------------------------------------------

The matter of assessing the quality of requirements has received significant attention, not only in the research literature   \cite{schneider_6.3.1_2000,soares_user_2011} but also in industry standards, from the venerable IEEE 830-1993 \cite{committee_ieee_1998} to the more recent ISO/IEC/IEEE 29148-2011 \cite{noauthor_iso/iec/ieee_2011}. They list such criteria as traceability, verifiability, consistency, justifiability and completeness.

For the present discussion, we need  criteria one notch higher in the abstraction scale since the goal is to evaluate not the quality of requirements but \textit{approaches} to requirements engineering. The discussion retains nine criteria for assessing requirements approaches.

Criterion 1, Audience, addresses the level of expertise expected of people who will use the requirements. Do they need, for example, to have received formal methods training? Or are the requirements intended for use by any stakeholder?

Criterion 2, Level of Abstraction (abbreviated in the assessment sections as ``\textit{Abstraction}''), addresses the level of detail (of properties of the system under description) which the requirements may or must cover. 

Criterion 3, Associated method (``\textit{Method}''), assesses whether the approach includes a comprehensive methodology to guide the requirements process (as opposed to more method-neutral approaches, which provide requirements support but adapt to their users' preferred methods). 

Criterion 4, Tool Support (``\textit{Tools}''), covers the availability of tools to support the approach, as opposed to approaches that are conceptual only.

Criterion 5, Traceability support (``\textit{Traceability}''), assesses how the approach handles one of the most important issues associated with requirements: keeping track of one- or two-way relations between requirement elements and their counterparts in design, code and other project artifacts. The IEEE standards emphasize the role of traceability as one of the key factors of requirements quality.

Criterion 6, Non-functional requirements (``\textit{Coverage}''), addresses whether the approach covers only the description of functional properties of the system (the functions it must perform, and environment constraints on these functions) or extends to non-functional properties (affecting aspects other than the system's function, such as performance and security).

Criterion 7, Environment vs System (``\textit{Scope}''), refers to the classic Jackson-Zave distinction \cite{Zave:1997:FDC:237432.237434} between two complementary parts of the requirements: describing the environment (or ''domain'') in which the future system will operate, and the constraints it imposes, such as ``no car will travel faster than 250 km/hour'' or ``any bank transfer above EUR 10,000 must be reported''; and the system (or ``machine'') which the project will build. Does the approach cover both, or only the system part? 

Criterion 8, Verifiability, assesses whether an approach supports the possibility of formally verifying properties of the resulting requirements. For approaches that provide such facilities, ``formal verification'' typically means mathematical proofs, preferably supported by tools since manual verification is not sufficient for large and complex systems.

Criterion 9, Semantic definition (``\textit{Semantics}''), assesses the availability and scope of a precise (if possible, formal) definition of the approach.

%----------------------------------------------------
\section{Review of important approaches}\label{sec:approaches}
%----------------------------------------------------

We now explore the approaches in the order of the five categories of the previous section, illustrating them through the example introduced in section \ref{sec:casestudy} and evaluating them per the criteria of section \ref{sec:sec:criteria}.

\subsection{Natural language}
\label{sec:sec:NL-based}

Natural language is, as noted, the dominant form of the requirements of  practical projects in industry. A number of requirements approaches consequently start from natural language statements of requirements. Such approaches  face a fundamental challenge: software construction needs a high degree of precision, but natural language is notoriously imprecise. (\cite{meyer_formalism_1993} is a detailed analysis of the problems of using natural language for requirements.) There are two ways of addressing these problems in practice:
\begin{itemize}
\item \textit{Analyze} the natural-language requirements texts, performing Natural Language Processing (``NLP'') to extract precise information and in particular to detect possible inconsistencies. Examples of methods using this approach include NL to OCL \cite{hahnle_authoring_2002}, NL to STD and NL to OWL.
\item \textit{Constrain} the kind of natural language used in requirements to ensure some degree of precision, without going as far as the semi-formal and formal approaches studied next. Examples include Requirements Grammar (\ref{sec:sec:sec:RequirementsGrammar}), Relax (section \ref{sec:sec:sec:Relax}) and Stimulus (section \ref{sec:sec:sec:Stimulus}).
\end{itemize}

%\textbf{I THINK THIS PARAGRAPH IS REDUNDANT, PLEASE CONFIRM.} Most of them propose to constrain the \gls{NL} to provide a more formal representation (see ???????????????????????????). We will present here approaches proposing a requirements dedicated language (Requirements Gramm\cite{scott_context-free_2004} \cite{whittle_relax:_2009} \cite{jeannet_debugging_2015}, and approaches that aims to transform from \gls{NL} to a formal representation \cite{hahnle_authoring_2002} \cite{aceituna_evaluating_2011} \cite{li_stakeholder_2015}.

\subsubsection{Requirements Grammar}
\label{sec:sec:sec:RequirementsGrammar}
% Do we conserve this approach?

Significant effort has been devoted to processing natural-language requirements automatically, with the purpose of detecting inconsistencies and, more generally, improving quality. Examples of such work include \cite{mich_nl-oops:_1996}, from 1996, and, more recently, \cite{slankas_automated_2013}. Full natural-language processing raises challenges at the frontier of artificial intelligence research. To make the task more tractable, the Requirements Grammar approach defines a structured, context-free subset of natural language. Then the idea is that requirements elicitation will produce requirements in this language, making it possible to avoid  inconsistencies. As with a programming language, the overall structure involves fixed keywords, borrowed here from English, such as \textbf{if} and \textbf{shall}, but they can be combined with free-form elements with no predefined meeting, as in

      \textbf{if} the gears are locked down, the doors \textbf{shall} be closed

\begin{figure}[htb]
\centering
\includegraphics[scale=0.28]{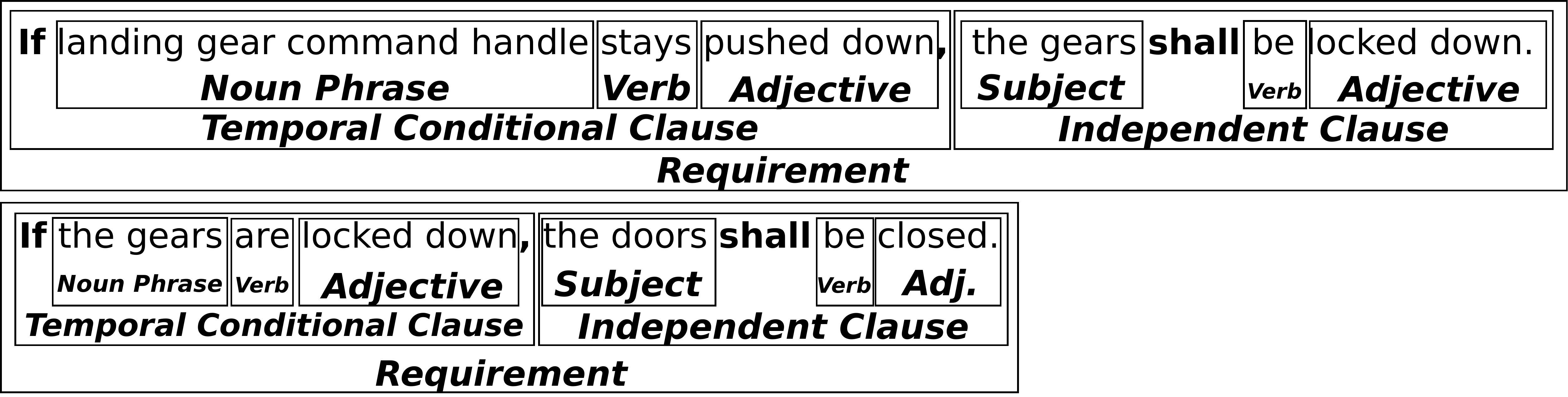}
\caption{Representation of requirement R11bis in Scott and Cook's Requirements Grammar}
\label{fig:RGExample}
\end{figure}

\myfig{RGExample} shows a representation of requirements R11bis (\emph{When the command line is working, if the landing gear command handle has been pushed DOWN and stays DOWN, then eventually the gears will be locked down and the doors will be seen closed.}) in this approach.  The boxes shows the hierarchical structure.
%------------------------
%------------------------

The grammar is not able to express this requirement as stated in \mysec{casestudy}:
each requirement can only involve one \emph{Independent Clause}, with a single \emph{Subject of interest}.
In the given requirement, there are two subjects (the \emph{gears} and the \emph{doors}). A solution is to split this requirement into two:
\begin{itemize}
\item When the command line is working, if the landing gear command handle has been pushed DOWN and stays DOWN, then eventually the gears will be locked down.
\item When the command line is working, if the gears are locked down, then eventually the doors will be seen closed.
\end{itemize}
Then we adapt the result to the grammar according to the following scheme, where keywords of the grammar appear in \textbf{boldface}: 
\begin{description}
\item[Requirement:] a sentence corresponding to a requirement.
	\begin{description}
	\item[Independent Clause:] the mandatory part of the requirement, that describe the need,
    	\begin{description}
    	\item[Subject:] the subject of interest of the requirement,
        \item[shall:] a keyword,
        \item[Verb:] the action of the requirement,
        \item[Adjective] or \textbf{Noun Phrase:} depending of the verb, a complement for the verb.
    	\end{description}
    \item[Temporal Conditional Clause:] an optional constraint on the requirement.
    	\begin{description}
        \item[If:] a keyword,
    	\item[Noun Phrase:] the subject of the constraint,
        \item[Verb:] the action of the constraint, 
        \item[Adjective] or \textbf{Noun Phrase:} depending of the verb, a complement for the verb.
    	\end{description}
	\end{description}
\end{description}

Assessing Requirements Grammar according to the criteria of section \ref{sec:sec:criteria}:
\begin{itemize}
\item \textit{Audience}: since the notation, while constrained, uses a subset of natural language, requirements are readable by any stakeholder, including those who are not aware of the constraints and will only see a natural-language description, possibly a bit contrived. 
\item \textit{Abstraction}: this approach is for requirements only, not influenced by implementation concerns.
\item \textit{Method}: the approach does not assume a particular requirements engineering method.
\item \textit{Tool}: the authors proposed a tool, named Badger\footnote{The tool seems no longer to be available.}, to express requirements and analyze their lexical clauses.
\item \textit{Traceability}: the approach focuses on requirements, independently of other steps and products (design, implementation), so it offers no specific support for traceability.
\item \textit{Coverage}: given that the approach expresses requirements in a very abstract form, it can include both functional and non-functional requirements.
\item \textit{Scope}: the approach can cover both system and environment aspects.
\item \textit{Verification} is limited to the following properties: consistency (the lexical clause are analyzed to detect if all requirements follow the pattern).
\item \textit{Semantics}: the syntax of specifications is defined precisely (through the concept of context-free Requirements Grammar) but there is no corresponding rigorous definition of the semantics. 
\end{itemize}

\subsubsection{Relax} \label{sec:sec:sec:Relax}
Relax \cite{whittle_relax:_2009} is a language for formal modeling of requirements of \gls{CAS}.
%Intended audience
The Relax syntax is close to natural language.
\myfig{NLLGS} shows requirements R11bis, R12bis, R21 and R22 expressed in it.

%------------------------
\begin{figure}[htb]
\begin{description}
	\item[R11bis:] The gear SHALL be locked down and the doors SHALL be closed AS EARLY AS POSSIBLE AFTER the landing gear command handle has been pushed down and stays down.
    \item[R12bis:] The gear SHALL be locked retracted and the doors SHALL be closed AS EARLY AS POSSIBLE AFTER the landing gear command handle has been pushed up and stays up. 
    \item[R21:] The retraction sequence SHALL not be observed AS EARLY AS POSSIBLE AFTER the command handle remains in down position.
    \item[R22:] The outgoing sequence SHALL not be observed AS EARLY AS POSSIBLE AFTER the command handle remains in up position.
\end{description}
\caption{Representation of Landing Gear System requirements expressed with Relax}
\label{fig:NLLGS}
\end{figure}
%------------------------
%Semantic definition
%Verifiability
Keywords such as \texttt{AS EARLY AS POSSIBLE} or \texttt{AFTER} express temporality of events. They are semantically defined through a fuzzy branching temporal logic (FBTL) \cite{moon_fuzzy_2004}, making it possible to submit the requirements to validation tools.
For example, R11bis can be translated into FBTL as:
\[AG (A\mathcal{X}_{> (landing\_gear\_command = down)_{d_2}} (A\mathcal{X}_{ \geq d_1} (AG (gear = locked\_down) \wedge AG (doors = closed))))\]
(``In any state after the event 'landing gear command as been locked down', 'gear is locked down and doors are closed' becomes true in a near future state.'') The example illustrates support for ``fuzzy'' notions (the F in FTBL) such as ``near''.

Relax places a particular emphasis on the expression of environment properties. One can express such properties directly through the keyword \verb$ENV$, define  ``monitors'' through \verb$MON$, and express relationships between them through \verb$REL$.

An original feature of Relax, explaining the name, is the ability to mark some requirements as critical, and to \textit{relax} a non-critical requirement if necessary to preserve the critical ones. 

Assessing Relax according to the criteria of section \ref{sec:sec:criteria}:
\begin{itemize}
\item \textit{Audience}: the notation uses natural language expressions, requirements are hence readable by any stakeholder. 
\item \textit{Abstraction}: this approach is for requirements only, not influenced by implementation concerns.
\item \textit{Method}: the approach does not assume a particular requirements engineering method.
\item \textit{Tool}: only propotypes such as Xtext \cite{eysholdt2010xtext} editors have been developed around Relax.
\item \textit{Traceability}: Relax focuses on requirements, independently of other steps and products (design, implementation), so it offers no specific support for traceability. Nevertheless the language provides a way to express relationships between requirements (through the keyword \verb$DEP$).
\item \textit{Coverage}: targeting adaptive systems, Relax mainly addresses functional requirements.
\item \textit{Scope}: the approach explicitly covers system and environment aspects.
\item \textit{Verification} is linked to the fuzzy branching temporal logic capabilities.
\item \textit{Semantics}: on the basis of a precisely defined syntax for specifications,the semantics is defined through FTBL (fuzzy branching temporal logic). 
\end{itemize}

\subsubsection{Stimulus}\label{sec:sec:sec:Stimulus}

%Tool support
%Level of abstraction
The Argosim Stimulus tool \cite{Stimulus} expresses requirements in a natural-language-like syntax \cite{jeannet_debugging_2015}, similar to Relax.
%Intended audience
%BM -- I removed the next statement, I think it brings nothing!
It is directed at stakeholders involved in system development.

\begin{figure}[htb]
\centering
\includegraphics[scale=0.5]{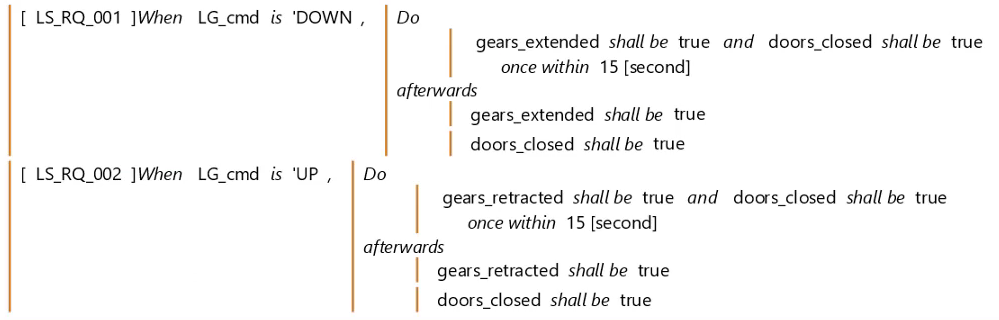}
\caption{A possible representation of requirements R11bis and R12bis in Stimulus}
\label{fig:StimulusExample}
\end{figure}

\myfig{StimulusExample} shows an initial attempt at expressing requirements \emph{R11bis} and \emph{R12bis}. 
It actually applies to a more complete version of the LGS example, taking into account timing properties from the original LGS paper not included above: a 15-second duration for retraction and for the outgoing sequence.

%------------------------
%------------------------

The idea is that after writing such an initial version one may, by simulating inputs and observing outputs, detect possible problems and improve the description. For example, Stimulus defines the semantic of \emph{When} as ``\emph{at the time the condition holds''}. 
Then \emph{R11bis} as defined in \myfig{StimulusExample} (requirement \emph{[LS\_RQ\_001]}) uses both a \emph{When} and a \emph{Do ... afterwards}.
Without this clause, the \emph{When} would mean that
When the handle is down, within 15 seconds the doors shall be closed and the gears down, leaving the behavior undefined afterwards. With the \emph{Do ... afterwards}, the meaning is that after the handle has been pushed down, within 15 seconds the doors shall be closed and the gears down, remaining so until a new event.

Such wrong behavior can be detected through the Stimulus model-checker, which simulates the system behavior by varying inputs.
Users can observe the system reaction and correct the requirements as needed. Stimulus favors such a process of incremental improvement of the requirements.

Assessing Stimulus according to the criteria of section \ref{sec:sec:criteria}:
\begin{itemize}
\item \textit{Audience}: the language is close to natural language; requirements are readable by any stakeholder. 
\item \textit{Abstraction}: this approach is for requirements only, not influenced by implementation concerns.
\item \textit{Method}: the approach does not assume a particular requirements engineering method.
\item \textit{Tool}: Stimulus is the name of both the language and the tool supporting it.
\item \textit{Traceability}: the approach focuses on requirements, independently of other steps and products (design, implementation), so it offers no specific support for traceability.
\item \textit{Coverage}: the approach covers only functional requirements.
\item \textit{Scope}: the approach explicitly covers both system and environment aspects.
\item \textit{Verification} can use model-checking to simulate a system's reaction to various inputs.
\item \textit{Semantics}: The language is inspired by Lucid Synchrone \cite{colaco_conservative_2005} and Lutin \cite{raymond_specifying_2008}, defined in the literature with precise semantics.
\end{itemize}

\subsubsection*{4.1.A NL to OCL}
\label{sec:sec:sec:NLtoOCL}

\cite{hahnle_authoring_2002} introduced a syntax to express constrained natural language specifications.
The tool is based on UML \cite{object_management_group_omg_uml_2015} and the associated OCL language \cite{object_management_group_omg_ocl_2014}, a Design-by-Contract-like mechanism for expressing semantic constraints on systems modeled in UML.
It formalizes constraints, originally expressed in constrained natural language, into \gls{OCL}. The long-term goal is integration into the KeY
Java-oriented formal verification project \cite{ahrendt2005key}. The presentation of NL to OCL described a tool for object-oriented modeling, intended to  allow non-experts to write constraints without having to embrace full-fledged formal methods.

\begin{figure}[ht!]
\begin{description}
  \small
  \item[Operation] \hspace{0.4cm} retract 
  \item[OCL:] \begin{spverbatim}
context LGS::retract()
pre: self.handle = up
post: self.doors = closed and self.gears = up

\end{spverbatim}
  \item[English:] \hspace{1ex} for the operation retract() of the
  class LGS, the following precondition should hold: \\
  \hspace*{2ex} the handle is up \\
  and the following post-conditions should hold: \\
  \hspace*{2ex} the doors are closed
  \hspace*{2ex} the gears are up.
\end{description}
\captionsetup{labelformat=empty}
\caption{\label{fig:ocl-eng}Fig.~1.A~Possible representation of requirement R12bis with the NL to OCL approach}
\vspace{-0.2cm}
\end{figure}

\hyperref[fig:ocl-eng]{Fig.~1.A} expresses requirement R12bis using the NL to OCL approach. This example (not tried out since the tool is no longer available) shows how to match an English specification of the \textit{retract} operation, including precondition and postcondition, with its representation in OCL.

Assessing tNL to OCL according to the criteria of section \ref{sec:sec:criteria}:
\begin{itemize}
\item \textit{Audience}: the target users of OCL are software engineers able to read a UML diagram and possessing a basic understanding of logic and formal reasoning.
\item \textit{Method}: the implied methodology starts from constrained natural-language requirements and expresses them in UML, with OCL for expressing semantic properties.
\item \textit{Tool}: the original tool for translation to OCL \cite{hahnle_authoring_2002} is no longer available. The UML part of the approach is supported by the wide range of available UML tools (including other approaches for verification such as \cite{Cabot:2007:UTF:1321631.1321737}).
\item \textit{Traceability}: the requirements are expressed as contracts on operations, linking them to the UML specification. There is, however, no specific support for traceability to software artifacts.
\item \textit{Coverage}: the approach applies to requirements whose semantics can be expressed through contracts.
\item \textit{Scope}: the approach only covers system aspects.
\item \textit{Verification}: the idea is to use the KeY tool to verify consistency of requirements.
\item \textit{Semantics}: the \gls{OCL} representation of requirements provides a semantic definition. 
\end{itemize}

%\subsubsection{NL to STD}
\subsubsection*{4.1.B NL to STD}
\label{sec:sec:sec:NLtoSTD}

The methodology of \cite{aceituna_evaluating_2011} iteratively transforms natural-language requirements into State Transition Diagrams, defined as sets triples [initial state, transition, resulting state].
%Non-functional requirements
%System vs environment

%------------------------
\begin{figure}[htb!]
\centering
\includegraphics[scale=0.1]{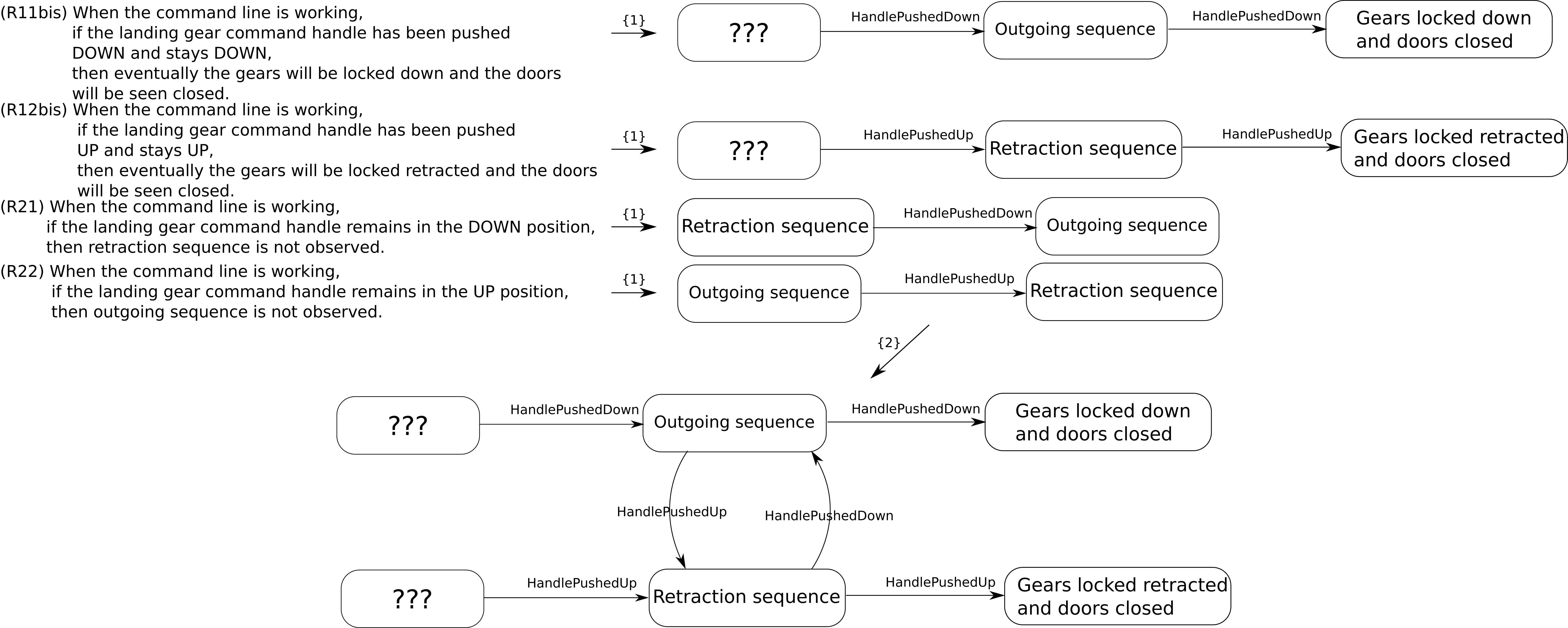}
%\caption{Transformation from NL requirements to STD}
\captionsetup{labelformat=empty}
\caption{Fig.~1.B~Transformation from NL requirements to STD}
\label{fig:NLtoSTD}
\end{figure}
%------------------------

\hyperref[NLtoSTD]{Fig.~1.B} illustrates the transformation of requirements into transition diagrams. The first step \{1\} transforms requirements into diagram fragments. For example, the requirement R11bis yields a transition from an unknown state, when the handle is down, to a final state \textit{Gears locked down and doors closed}. Such states will have to be refined later. Step \{2\} assembles these fragment into a single diagram, representing the entire system but not final since it may still contain unknown states. Step \{2\} adds missing information and repeats the process.

Assessing the NL to STD approach according to the criteria of section \ref{sec:sec:criteria}:
\begin{itemize}
\item \textit{Audience}: the requirements are first expressed in natural language, readable by all stakeholders. The transition diagrams resulting from the transformation in step 2 are meant for a more expert audience, but they include natural-language explanations coming from the original text.
\item \textit{Abstraction}: this approach is for requirements only, not influenced by implementation concerns.
\item \textit{Method}: the approach proposes a method as outlined (going from natural language to transition diagrams).
\item \textit{Tools}: the translation into transition diagrams is manual. 
\item \textit{Traceability}: there is no specific support for traceability.
\item \textit{Coverage}: given that the approach expresses requirements in a very abstract form, it can include both functional and non-functional requirements.
\item \textit{Scope}: the approach can cover both system and environment aspects.
\item \textit{Verification}: transition diagrams are formal texts, which can be submitted to tools.
\item \textit{Semantics}: State Transition Diagrams are a well-known notion with precise semantics.
\end{itemize}

\subsubsection{NL to OWL}\label{sec:sec:sec:NLtoOWL}
%Associated method
The approach of \cite{li_stakeholder_2015} translates natural-language requirements into an intermediate requirements modeling language that can be easily formalized in OWL \cite{bechhofer_owl:_2009}.
%Intended audience

In the example, requirement R11bis -- \emph{When the command line is working, if the landing gear command handle has been pushed down and stays down, then eventually the gears will be locked down and the doors will be seen closed} -- yields two functional goals:
\begin{quote}
\it FG11-1 := lock\_down $<$object: \{the gears\}$>$ :$<$ $<$trigger: push\_down $<$object: \{landing gear command handle\}$>$$>$ \\
FG11-2 := close $<$object: \{the doors\}$>$ :$<$ $<$trigger: FG11-1$>$$>$
\end{quote}
FG11-1 states that the gears should be locked down a while after the landing gear command handle has been pushed down. FG11-2 states the obligation to close the doors when the gears are locked down, as triggered by the first functional goal. R12bis can be modeled in a similar way.

The requirement R21 -- \emph{When  the  command  line is  working, if the landing gear command handle remains in the down position, then retraction sequence is not observed} -- can be modeled using only one functional goal:
\begin{quote}
\it FG5 := extend $<$object: \{the gears\}$>$ :$<$ $<$trigger: remains\_down $<$object: \{landing gear command handle\}$>$$>$
\end{quote}
This functional goal models the need to do not observe retraction sequence -- and so, extend the landing gears -- when the handle remains down.
Requirement R22 can be modeled in a similar way.

Assessing NL to OWL according to the criteria of section \ref{sec:sec:criteria}:
\begin{itemize}
\item \textit{Audience}: requirements engineers.
\item \textit{Abstraction}: requirements only, no tainting by implementation.
\item \textit{Method}: NL to OWL guides the requirements process, from natural language to more formal.
\item \textit{Tool}: none directly for the method; there are tools for OWL such as Protégé \cite{Protege}.
\item \textit{Traceability}: requirements are decomposed into ontologies, sharing the same namespace, which can be used to create links between several requirements.
\item \textit{Coverage}: both functional  (\emph{functional goals}) and non-functional properties (\emph{quality goals}).
\item \textit{Scope}:  both system and environment (represented through \emph{domain assumptions}).
\item \textit{Verification}: no support for requirements verification of  (though OWL has a formal semantics).
\item \textit{Semantics}: from OWL.
\end{itemize}

\subsection{Semi-formal}
\label{sec:sec:sf_based}
%\jmb{I believe "General purpose" than "Commercial" more suitable for the all section (SysML is not commercial for example).}

A number of approaches, including both research efforts and industrial products, use partially formalized notations. Some of the most important, reviewed below, are Doors, Reqtify, KAOS and SysML, as well as (in the full version only) URN and URML.
%\cite{gea_requirements_2011}. 

\subsubsection{Doors and Reqtify} \label{sec:sec:sec:DoorsReqtify} 

Doors \cite{DOORS} and Reqtify \cite{Reqtify} are widely used in industry. They are semi-formal in the sense that they require a partially structured approach to requirements management. While they are distinct products from different providers (respectively IBM Rational and Dassault Systems), they are often used jointly and we cover them together. In both cases the focus is not on producing requirements, but on managing requirements independently of how they were produced.

Doors is a collaborative tool allowing different stakeholders to work on requirements, typically maintained as spreadsheets, and set priorities according to levels of risks. Reqtify's focus is on traceability: the tool supports defining relationships between requirements typically expressed in natural language and coming from such tools as Microsoft Word, spreadsheets or modeling tools.

Since these approaches do not define any specific method or notation for expressing requirements, we cannot demonstrate them on the running case study. In the operational practice the requirements would be expressed in some document, e.g. Word or PDF. DOORS would then record this document and its various attributes (creation date, version number, priority...) in its database, supporting the management of these requirements throughout the project.  Reqtify would support defining and managing a traceability between their various elements. 

With respect to the criteria of section 2: 
\begin{itemize}
%Intended audience – prerequisites - N none
\item \textit{Audience}:
aimed at a large audience of stakeholders, no particular technical prerequisite.
\item \textit{Abstraction}:
%Level of abstraction - Low
focused on requirements  no influence from implementation.
\item \textit{Method}:
%Associated method (N) 
these are software tools, with no particular method attached.
\item \textit{Traceability}:
traceability is the strong point of Reqtify in particular, which offers support for tracing requirements from specification to design and code. For example, it makes it possible to import requirements expressed in a Microsoft Word document and link them to C code. 
\item \textit{Coverage}:
%Non-functional requirements (Y)
%System vs environment - B both of them
no particular restriction of coverage or scope.
\item \textit{Semantics}:
nor formal semantics, no verification methodology.
\end{itemize}

\subsubsection{KAOS} \label{sec:sec:sec:kaos}
KAOS \cite{DARDENNE19933}, like i* (\cite{yu1997towards}), is based on the Goal-Oriented Requirements Engineering approach to requirements \cite{lamsweerde_goal-oriented_2001} \cite{lamsweerde:goals-keynote}  \cite{mylopooulos:goals}. The key idea of this approach is to base requirements on a higher-level concept, goals. A goal is statement of intent expressed in terms of business needs (such as ``turn more sales inquiries into actual sales'' for a customer management sytem). Requirements then express system properties helping to achieve these goals. (Specifically, R, A, D $\models$ G with sets of requirements R, domain assumptions A, domain properties D and goals G.) Goals can be composite, expressed in terms of simpler goals through trees operators including AND, OR and ``+'' (denoting a less formal relation, ``contributes to''). The general approach is refinement-based: start from high-level goals and decompose them using the operators. A
 non-composite goal is called a ``requisite''. The OR operator makes it possible to include alternative paths.

KAOS uses natural language to express goals and a semi-formal notation for relationships between goals, with concepts such as ``milestone'' and `` conflict'', and supports the refinement process.  

Goals cover both system and environment properties: if a requisite can be assigned to an agent of the system, it is an ``operational goal'', describing a system property. Otherwise it is an ``expectation'', describing an environment property. 

%-------------
\begin{figure}[htb]
\includegraphics[scale=0.4]{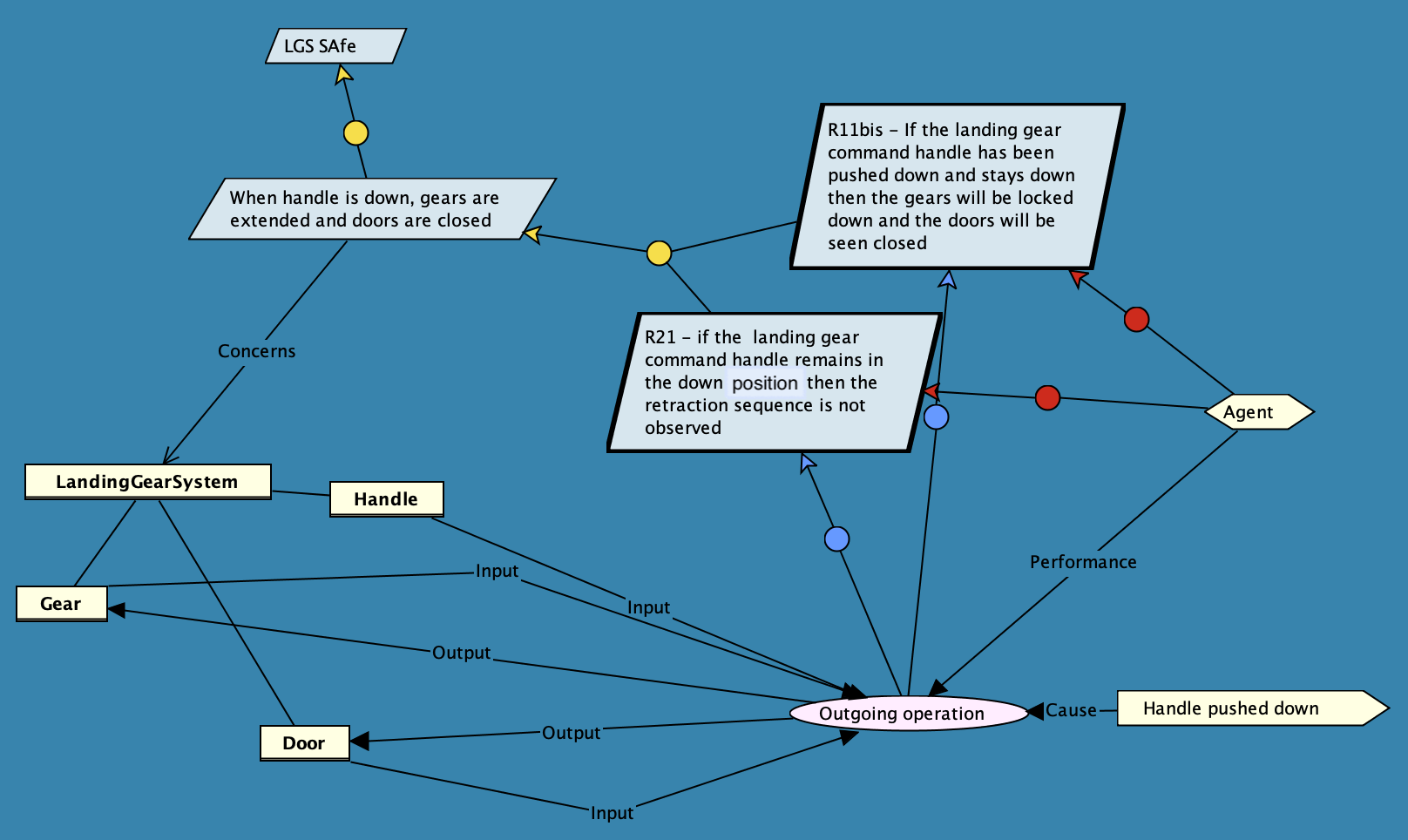}
\caption{Excerpt of the KAOS diagram of requirements R11 bis and R21 of the LGS (\textit{Objectiver}) } 
\label{fig:LGSKaos}
\end{figure}
%-------------

The LGS model of \myfig{LGSKaos} covers entities including door, gear and handle. Both  R11bis  and R21 refine the goal «When handle is down, gears are extended and doors are closed», which assumes normal mode and is itself part of the refinement of a more global goal defining the whole system's safety. «Outgoing operation» addresses R11bis and R21 by managing the LGS outgoing sequence: after an the LG command, the handle has been moved up,  doors remain closed and gears locked down. Some agent, triggered by the event ``handle pushed down'', will be responsible for this operation.

Assessing KAOS according to the criteria of section 2:

\begin{itemize}
\item \textit{Audience}:
%Intended audience – prerequisites (S: specific training required)
KAOS requires modeling experience, and some training in the method.
\item\textit{Abstraction}: 
no influence from implementation. KAOS can in fact be described as particularly abstract since it focuses on a concept, goals, which is even higher than requirements.
%Level of abstraction (Both)
\item\textit{Method}: 
KAOS includes a general methodology for modeling systems, specifying dependencies between requirements, and refining goals.
%Associated method (Y)
\item\textit{Tool}: 
%Tool support (Y)
Objectiver \cite{Objectiver} supports the expression and refinement of user requirementsin KAOS. 
\item\textit{Traceability}: 
%Traceability support (Y)
there is support for linking to specification documents.
\item\textit{Coverage}: 
%Non-functional requirements (Y)
mostly functional requirements, but can include some non-functional ones.

\item\textit{Scope}: 
%System vs environment (Both)
both system and environment through the notion of operational goal and expectation in the refinement process, as described above.
\item\textit{Verification}: 
%Verifiability (N) 
no specific support (graphical notation).
\item\textit{Semantics}: 
%Semantic definition (Y)
behavioral goals can be described in temporal logic (LTL) or in event-B (\cite{Abrial2010}) \cite{Matoussi2010AnEF}. 
\end{itemize}

%URN ==============================
\subsubsection*{4.2.A. URN}
\label{sec:sec:sec:urn} 
User Requirements Notation \cite{Amyot2003}  is a recommendation of the International Telecommunication Union (standard Z.151, from 2008, updated 2012, third version in progress)  for the modelling, analysis, specification and validation of requirements, used mainly for business process modeling.

The basic concepts are  goals, scenarios, and links between such elements.
URN combines two complementary views: static goals, through the Goal-oriented Requirement Language (GRL); and dynamic scenarios, through the Use Case Map (UCM) notation. Both notations support checking: for GRL models, through “strategies”, representing initial situations; for UCM models, through “scenarios”, similar to test cases. There is also a framework for formal verification of UCM  (\cite{10.1007/978-3-642-11811-1_4}, including time extensions. 
For traceability:
\begin{itemize}
\item 
Internally: URN supports links connecting GRL and UCM models, enabling completeness and consistency analysis. 
\item
With external notations: through such tools as jUCMNav \cite{Jucmnav}, to integrate URN models with DOORS.

\end{itemize}
For verification, methodological elements support validating goal-oriented models and resolving conflicts (\cite{Hassine:2016:QSM:2944570.2944591} \cite{Hassine:2017:EAT:3054641.3054669}).

\begin{figure}[htb]
\includegraphics[scale=0.4]{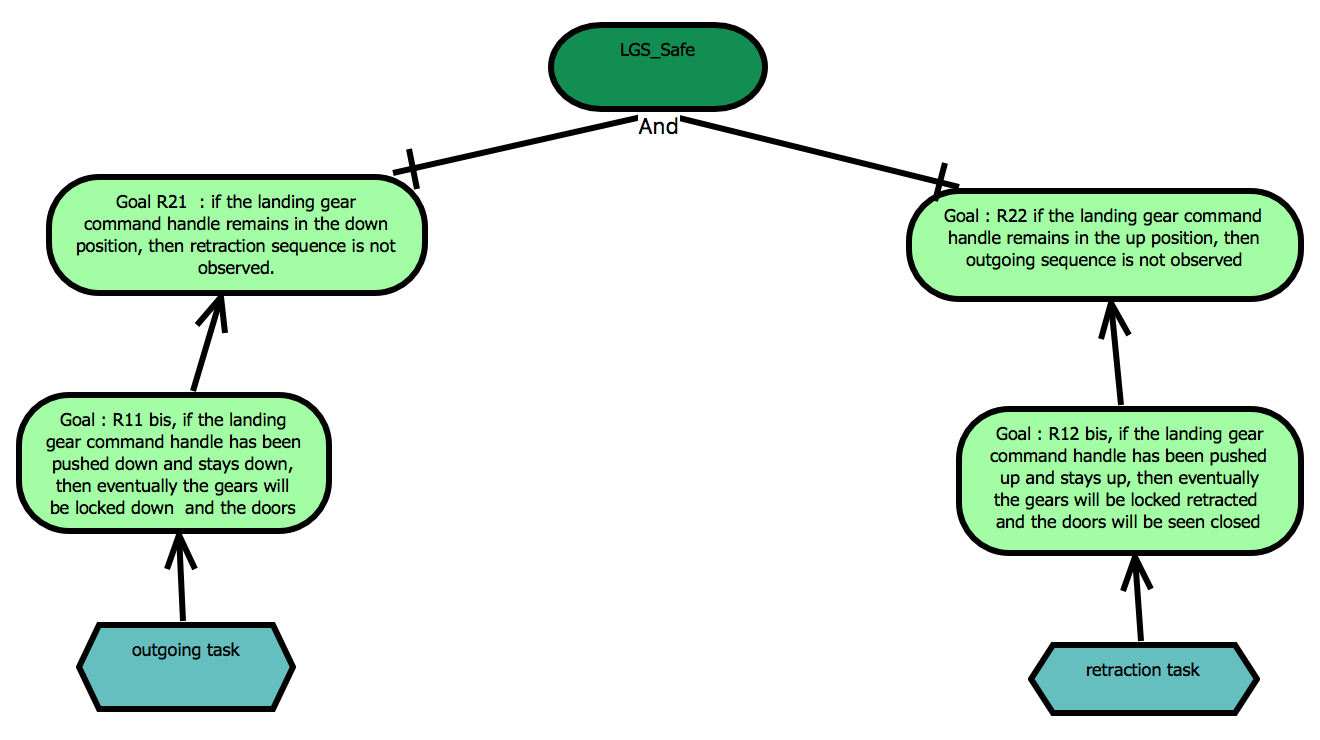}
\captionsetup{labelformat=empty}
\caption{Fig.~2.A~Partial URN diagram for LGS requirements R11 bis and R21 (\textit{jUCMNav})} 
\label{fig:LGSURN}
\end{figure}

Figure~\hyperref[fig:LGSURN]{2.A} uses GRL to describe the goals of the LGS example. The overall goal \textit{LGS Safe} is decomposed into two goals (among others), reflecting R21 and R22 and refined further into more specific goals reflecting R11bis and R12bis. A task contributes to a  goal: the ``outgoing'' task contributes to R11bis and ``retracting'' to R12bis. Note that tasks may depend on resources (that is not the case here). 
%-------------
%-------------

Assessing URN according to the criteria of section 2:
\begin{itemize}
\item \textit{Audience}:
%Intended audience – prerequisites (S: specific training required)
URN needs a specific training on GRL and UCM. % (the focus on requirements can  be done through GRL only).
\item \textit{Abstraction}: 
%Level of abstraction (H)
the approach does not have implementation concerns
\item \textit{Method}: 
%Associated method (Y)
URN does not impose any development process, but tutorials about jUCMnav present methodological elements.
\item \textit{Tools}: 
%Tool support (Y)
jUCMNav is the open-source Eclipse \cite{Eclipse} plugin that supports URN (\cite{roy_towards_2006}). Others include OpenOME \cite{OpenOme} %(a general, goal-oriented and agent-oriented modeling and analysis tool that supports GRL), 
Sandrila \cite{Sandrila} %(a commercial set of stencils for Visio - Sandrila SDL supports the URN notation), 
UniqueSoft \cite{Uniquesoft} %(has got a commercial tool that embeds jUCMNav in a way that allows formal specifications modeled with UCM), 
ArchSync \cite{ArchSync} %(a UCM tool that helps architects to reconcile a scenario-based architectural description with its source code, as changes are being made on the code) 
and TouchCORE \cite{TouchCORE} %(a support of GRL in the context of aspect-oriented modeling) 
\item \textit{Traceability}: 
%Traceability support (Y)
as noted, URN support links between model elements.
\item \textit{Coverage}:  
%Non-functional requirements (Y)
GRL focuses on requirements, especially non-functional ones. UCM is most useful for specifying functional requirements.  
\item \textit{Scope}: 
%System vs environment (S)
URN is dedicated to systems and specifically to reactive systems and business systems.
\item \textit{Verification}: 
%Verifiability (N) 
no formal support (beyond the methodological elements mentioned above).
\item \textit{Semantics}:
%Semantic definition (N)
from \cite{DBLP:journals/jsw/AmyotM11}, “\textit{the URN standard describes the URN abstract and concrete syntaxes formally, together with well- formedness constraints. However, the semantics is currently described more informally}”.  \end{itemize}
%URN ==============================

\subsubsection{SysML (Systems Modeling Language)} 
\label{sec:sec:sec:sysml}
\begin{figure}[htb]
\includegraphics[scale=0.4]{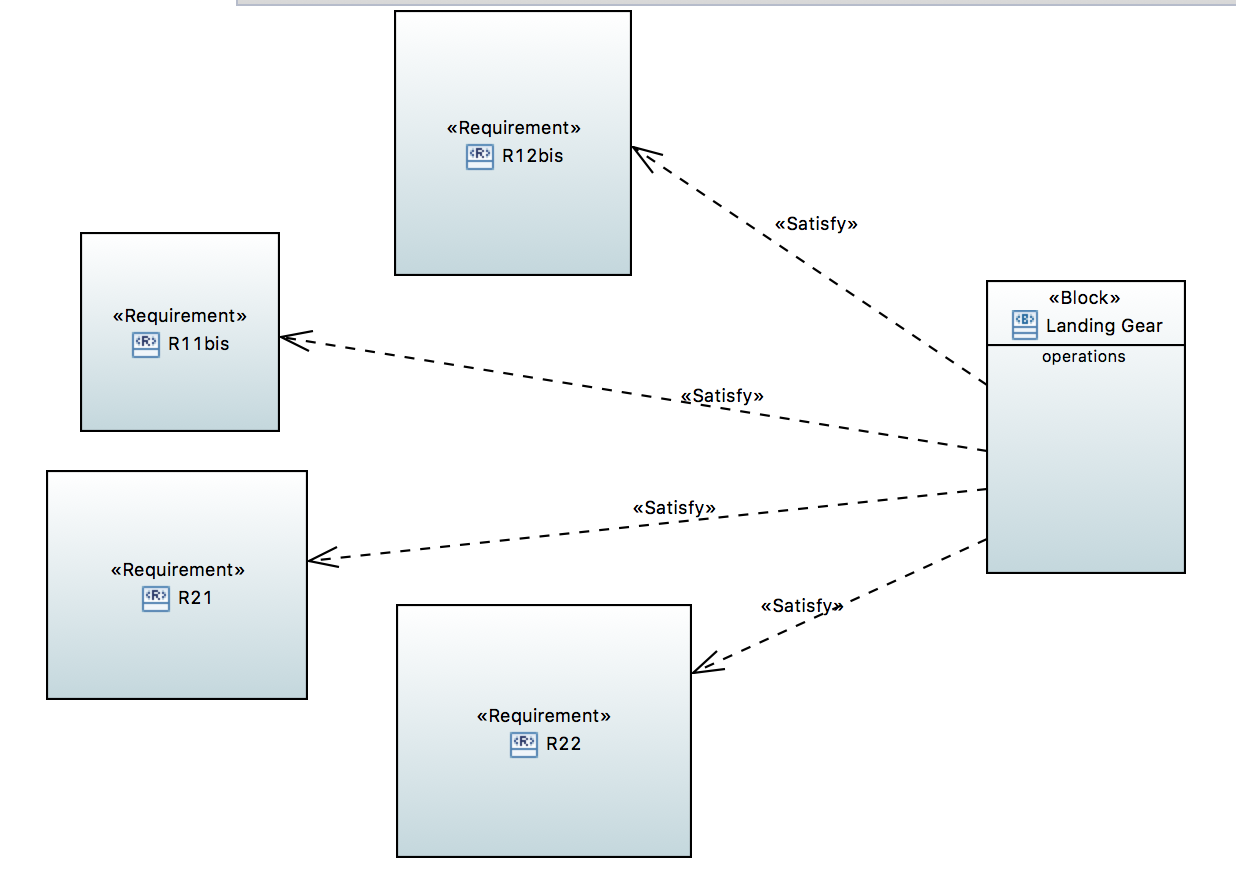}
\caption{Exerpt of the SysML functional requirements diagram of LGS (\textit{Papyrus}) } 
\label{fig:LGSReqDiag}
\end{figure}

SYStem Modelling Language  \cite{omg_omg_2007} is an extension of UML \cite{object_management_group_omg_uml_2015} dedicated to systems engineering. 
SysML provides requirements diagrams, which allow users to express requirements in textual representation, and cover non-functional requirements. 
The requirements diagram of SysML provides ways to express traceability links between different requirements (containment, derive, copy, trace) or between requirements and implementation elements (satisfy, verify, refine). 
Relationships between requirements and other modeling artifacts (like blocks, use cases, activities, \ldots), allow basic verification (e.g.  each requirement is supported by at least one modeling element, \ldots). 
A formal expression of requirements and thus their formal verification is not possible.
The SysML requirements diagram of \myfig{LGSReqDiag} expresses the links between the LGS requirements R11bis to R22 and the landing gear block.
%, as described in the block definition diagram of table \myfig{LGSReqTable}). 
%-------------
%-------------
%-------------
%\begin{figure}[htb]
%\includegraphics[scale=0.36]{img/ReqTable}
%\caption{Excerpt from the description of LGS'requirements (\textit{Papyrus}) } 
%\label{fig:LGSReqTable}
%\end{figure}
%-------------

Assessing SysML according to the criteria of section 2:
\begin{itemize}
\item \textit{Audience}:
%Intended audience – prerequisites (S: specific training required)
SysML is a modeling language needing specifc knowledge. 
\item \textit{Abstraction}: 
%Level of abstraction (H)	
SysMl stands at a high level of abstraction. Its requirement diagram enables quick requirements analysis and visual design.
\item \textit{Method}: 
%Associated method (N)
SysML, like UML, is a notation, providing no methodology.  
\item \textit{Tool}: 
%Tool support (Y)
It is supported by a number of tools such as IBM Rhapsody \cite{Rhapsody}, Modelio \cite{Modelio}, Enterprise Architect \cite{EnterpriseArchitect}, Papyrus \cite{Papyrus}, \ldots that implement a methodology preconised for using SysML. 
\item \textit{Traceability}: 
%Traceability support (Y)
SysML provides traceability links.
\item \textit{Coverage}: 
%Non-functional requirements (Y)
SysML supports both functional and non-functional requirements. 
\item \textit{Scope}: 
%System vs environment (S only System)
SysML focuses on system requirements, particularly for complex systems.
\item \textit{Verification}: 
%Verifiability (N) 
SysML does not allow any verification.
\item \textit{Semantics}: 
%Semantic definition (N) 
SysML does not provide any semantic definition of the approach. 
\end{itemize}

%URML ==============================
\subsubsection*{4.2.B. URML}
\label{sec:sec:sec:urml}
User Requirements Modeling Language \cite{helming_towards_2010, berenbach_use_2012} is a UML profile developed in collaboration between the Technical University of Munich and Siemens.  The approach unifies concepts from goal-oriented, feature-oriented, process-oriented, and risk-oriented approaches, integrating them into models covering both functional and nonfunctional requirements.

RML is a language in its own right, based on the ``Meta-Object Facility'' (MOF), with a meta-model mapped to a UML profile to implement the supporting tools. As a consequence, URML provides a graphical, icon-based notation to express requirements as well as associated notions such as threats, hazards, mitigations and even product lines and stakeholders. 

It is possible to associate semantic properties --- such as presupposes, details, constrains, refines, with, … --- with URML links between requirements and other artifacts.

The modeling focus is on systems, but URML can also describe properties of the environment. 

URML is more particularly dedicated to requirements elicitation, and does not provide any specific methodology to construct models, but it can support any methodology compatible with the URML meta-model. 

\begin{figure}[htb]
\includegraphics[scale=0.4]{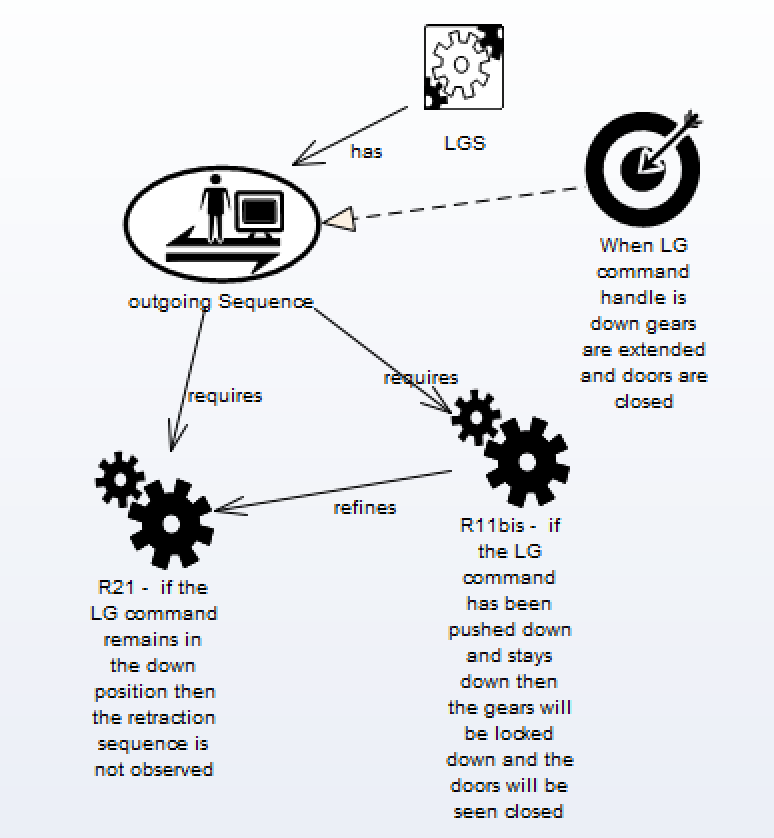}
\captionsetup{labelformat=empty}
\caption{Fig.~2.B~Excerpt of the URML requirements diagram of LGS (\textit{Enterprise Architect}) } 
\label{fig:LGS_URML}
\end{figure}

\hyperref[fig:LGS_URML]{Figure~2.B} is an excerpt of a URML model of the LGS. To ensure a basic functionality of the LGS and its safety, we need to consider the outgoing sequence of the gears. This process is supposed to achieve the goal ``when the LG command is down, gears are extended and doors are closed''. To do so, it requires both R21 and R11bis. R11bis is considered a refinement of R21 since it expresses the same need at a finer level of detail. 

%-------------
%-------------

Assessing URML according to the criteria of section 2:
\begin{itemize}
\item \textit{Audience}:
%Intended audience – prerequisites (S: specific training require)
as URML is an icon-based graphical modeling language, the learning process of abstractions of this requirements language will be easy. 
\item \textit{Abstraction}: 
%Level of abstraction (H)
URML is based on abstractions of requirements.   
\item \textit{Method}: 
%Associated method (N)
this approach does not assume a particular requirements engineering method
\item \textit{Tool}: 
%\item \textit{Semantics}: %Tool support (Y)
it is supported by an Enterprise Architect Add-On. \cite{EnterpriseArchitect}. 
\item  \textit{Traceability}: 
%Traceability support (Y)
the links between concepts can lead to traceability. 
\item \textit{Coverage}: 
%Non-functional requirements (Y)
URML integrates functional and nonfunctional requirements. 
\item \textit{Scope}: 
%System vs environment (B Both)
it is focused on systems and particularly complex systems
\item \textit{Verification}: 
%Verifiability (N) 
no mean is provided to do verification of requirements. 
\item \textit{Semantics}: 
%Semantic definition (Y)
some semantics can be put on links between requirements and other artifacts. 
\end{itemize}
%URML ==============================

\subsection{Graphs and automata}
%\TODO{JMB}
%\TODO{Sophie - review text - Done}
A number of approaches rely on the mathematical theories of graphs and automata, well known in computer science \cite{Minsky:1967:CFI:1095587} and supported by effective graphical representations.

Such approaches are often used to represent the dynamic aspects of a system, particularly behavior and timing. 
We examine
Petri nets,
Statecharts, 
Problem Frames, 
FSP/LTSA,
and FORM-L.
Other approaches offering different flavors of the same concepts include UML Activity Diagrams and State Diagrams \cite{omg2011umls}, UPPAAL \cite{uppaal2k}, DEVS \cite{devs} and SCXML \cite{w3c_scxml}.

%===========PetriNets====================
\subsubsection*{4.3.A Petri Nets}
\label{sec:sec:sec:petrinets}

\begin{figure}[htb]
\includegraphics[scale=0.55]{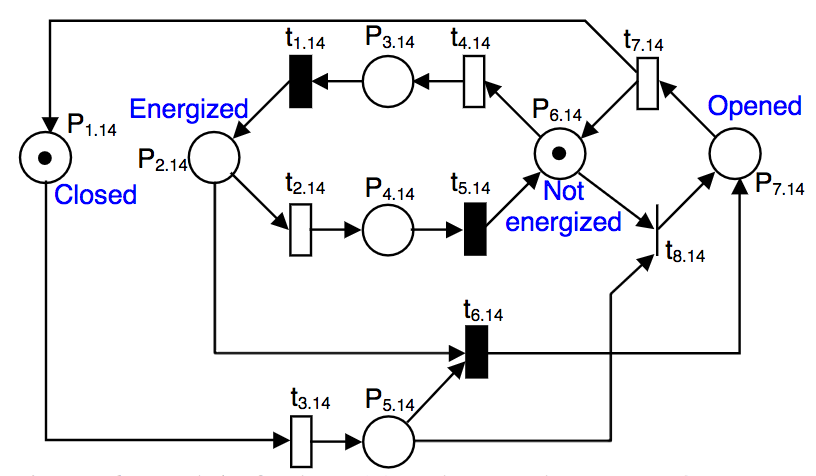}
\captionsetup{labelformat=empty}
\caption{Fig.~3.A~Petri Net model of an Electrical Circuit of Negative Pressuring Electro-valve (taken from \cite{Villani04.2})}
\label{fig:LGS-petri}
\end{figure}

A Petri net \cite{petrinets} is a directed graph, where each node is either a place (circle, representing a condition) or a transition (bar) and each edge is associated with a transition. An example appears in \hyperref[fig:LGS-petri]{Figure~3.A}.

Petri nets have been known for several decades and have been used for many applications, Their attractiveness comes from the simplicity of the model and its ability to describe processes in a clear, visual way with a precise mathematical basis. They suffer, however, from a lack of compositionality making them unsuitable, in the view of critics, for scaling up to the description of large systems. A good analysis of the pros and cons can be found in\cite{ZurawskiZ94}. 

%------------------------
%------------------------

Assessing Petri nets according to the criteria of section \ref{sec:sec:criteria}:
\begin{itemize}
\item \textit{Audience}: the notation, while graphical and clear, has to be learned. 
\item \textit{Abstraction}: this approach is for requirements only, not influenced by implementation concerns.
\item \textit{Method}: the approach does not assume a particular requirements engineering method.
\item \textit{Tool}: Petri nets have been around for a long time; many tools have been developed to edit, execute and verify nets.
\item \textit{Traceability}: the approach focuses on the expression of dynamic behavior requirements, and offers no specific support for traceability.
\item \textit{Coverage}: Petri nets in their basic form address functional requirements.
\item \textit{Scope}: Petri nets only cover the modeling of the precise behavior of some parts of the system rather than the environment.
Is is hence limited to fully express requirements of a system.
\item \textit{Verification} is supported by Petri-net tools, relying on model-checking.
\item \textit{Semantics}: Petri nets have an exact mathematical definition of their execution semantics, with a well-developed mathematical theory for process analysis. 
\end{itemize}
%=============PetriNets===============

\subsubsection{Finite automata, state diagrams and statecharts} \label{sec:sec:sec:statecharts}

    The most widely used kind of automaton, notable for the simplicity and power of the concept, is the \textit{finite} automaton, used (in applications to system modeling) through the closely related mechanism of a  \textit{finite-state diagram}. Such a diagram is a mathematical device defined by a finite set of \textit{states}, each representing a possible configuration of a system or computation, the designation of some of the states as \textit{initial} and some as \textit{final}, and a finite set of \textit{transitions} between states. Each transition models the effect of a given event in a certain state, by defining the resulting state or states. To model an entire system execution, the automaton starts in an initial state then processes events by following the corresponding transitions until it reaches a final state.
    
\begin{figure}[htb]
    \includegraphics[scale=0.5]{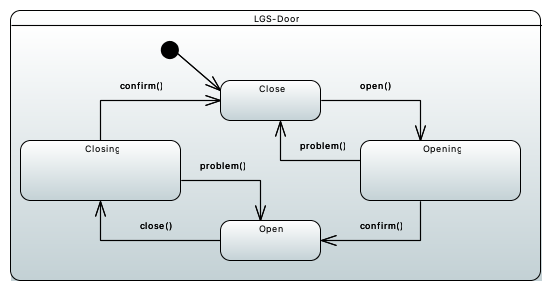}
    \caption{A finite-state diagram for the LGS Door}
    \label{fig:LGSSD}
\end{figure}

    Part of the attraction of finite-state diagrams is that they enjoy a natural representation as graphs, with nodes as states and transitions as edges, as illustrated in \myfig{LGSSD}. 

%------------------------
%------------------------

The most widely used finite-state diagram notation for describing the behavior of systems is the Statechart \cite{Harel87} or ``Harel chart''. 
A system is described through a finite number of states and transitions between those states (see \myfig{LGSSD}). Statecharts enjoy a well-documented formal definition and are suitable for the modeling of parallel systems.

Assessing Statecharts according to the criteria of section \ref{sec:sec:criteria}:
\begin{itemize}
\item \textit{Audience}: the notation, while graphical, assumes an understanding of the semantics. 
\item \textit{Abstraction}: Statecharts are meant for specification, independently of implementation concerns, and is abstract due to its specific notation. Note that available tools provide an execution environment to animate state machines (for simulation rather than actual implementation).
\item \textit{Method}: the approach enforces a strong methodological discipline, based on modeling systems in the form of states and transitions.
\item \textit{Tool}: many tools and languages are using Statecharts principles, sometimes redefining some of the semantics (e.g., UML). 
\item \textit{Traceability}: the approach focuses on the expression of dynamic behavior requirements and offers no specific support for traceability.
\item \textit{Coverage}: Statecharts mainly address functional requirements. 
\item \textit{Scope}: there is no specific modeling of the environment, although events leading to transitions can come from the outside or the inside of the system.
\item \textit{Verification}: supported by the semantics implementation of the tools, using model-checking.
\item \textit{Semantics}: the basic semantics of Statecharts comes from the theory of finite-state automata. 
\end{itemize}

\subsubsection{Problem Frames} \label{sec:sec:sec:problem_frames}
\begin{figure}[hbt]
    \includegraphics[scale=0.55]{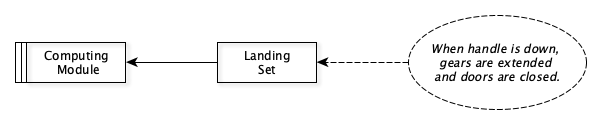}
    \caption{LGS Landing Set Problem Diagram}
    \label{fig:LGS-PF}
\end{figure}

\cite{Jackson:2000} is an approach to software requirements analysis developed by Michael Jackson in the nineties.
It has influenced number of subsequent approaches because it brought to light the need to divide requirements into \textit{system} and \textit{environment} properties. 
As defined by M. Jackson, a \textit{problem frame} ``defines the shape of a problem by capturing the characteristics and interconnections of the parts of the world it is concerned with, and the concerns and difficulties that are likely to arise.''
Problem Frames provide a methodology for decomposing requirements, treated as relationships between the system and the real world.
 \myfig{LGS-PF} depicts a problem diagram with a concrete \textit{machine domain} on the left (\texttt{Computing Module}), its corresponding \textit{domain} in the middle (\texttt{Landing Set}) and the \textit{requirement} on the right that lead to this domain.
The Computing Module is the software control machine that controls the Landing Set domain to ensure the behavioral requirement mentioned on the right.
By dividing a software purpose into a set of manageable and well documented pieces it is easier to apprehend a complex problem.
In addition this decomposition makes the pieces easier to reuse in the sense that when reused, one can benefit from the exact context (domain and requirement) of this particular piece of software.

\vspace{-0.2cm}
%------------------------
%------------------------  
\vspace{-0.2cm}

The Problem Frames approach is important for its historical influence on other approaches and its clear distinction between system and  environment. Assessing it accor    ding to the criteria of section \ref{sec:sec:criteria}:
\begin{itemize}
\item \textit{Audience}: The notation, while graphical and intuitive, is very specific and is targeted to Problem Frames specialists only. 
\item \textit{Abstraction}: this approach is for requirements mostly, not influenced by implementation concerns, and is very abstract due to its specific notation.
\item \textit{Method}: the approach is strongly related to  Jackson's methodology but has no particular traceability nor non-functional requirements support. 
\item \textit{Tool}: the approach provides a set of notations, some of them graphical
such as context diagrams or problem diagrams
(see \myfig{LGS-PF}). 
\item \textit{Traceability}: the approach focuses on the expression of dynamic behavior requirements and offers no specific support for traceability.
\item \textit{Coverage}: Problem Frames mainly address functional requirements.
\item \textit{Scope}: Problem Frames focus on the modeling of the system in its environment, treating both aspects as equally important.
\item \textit{Verification}: the approach is not designed for verification.
\item \textit{Semantics}: while precise, the graphical notation of Problem Frames has no precise semantics. 
\end{itemize}

%Verifiability

\subsubsection{FSP/LTSA} \label{sec:sec:sec:ltsa}
% http://www.doc.ic.ac.uk/ltsa/

``Process algebras'', which provide a formal basis for describing interactions between processes, have influenced several methods. The original process algebras were Hoare's CSP (Communicating Sequential Processes) \cite{Hoare:1978} and Milner's CCS (Communicating Sequential Processes), later extended to  cover mobile agents in the $\pi$-calculus \cite{Milner:1982}. FSP (Finite State Processes) proceeds from both CSP and CCS.

%The basic unit of all process algebras is a communicating concurrent process with its input and output ports, which makes feasible modeling the interaction with the environment and expressing requirements beyond the system itself. FSP not only offers mechanisms to describe processes’ behaviors but also supports expression and verification of liveness, safety, progress, fairness, and temporal properties through the powerful tool that supports FSP — LTSA.

The basic unit of all process algebras is a concurrent process which can communicate with others through input and output ports. For requirements, communication can model interaction both within the system and with the environment. Specifications in this style can formally express such essential temporal properties as liveness, safety, progress and fairness, and use supporting tools to verify them.

For FSP the supporting tool is LTSA (Labeled Transition System Analyzer \cite{LTSA}). From an FSP model, LTSA generates a  Labeled Transition System (LTS), suitable for automated analysis and animation. LTSA is \textit{compositional}, meaning that it is possible  to model the components of a system separately then, with process calculus mechanisms, their composition. In addition to helping the modeling process, LTSA's compositionality benefits the verification process: one may verify safety and liveness properties by model-checking individual components then their composition.

FSP can take advantage of compositionality to model the LGS example as the parallel composition (expressed through the \verb$||$ operator) of two distinct processes:

{\footnotesize\begin{verbatim}
||LGS = (LGS_BEHAVIOR || CONTROL_HANDLE)
\end{verbatim}}
CONTROL\_HANDLE specifies how the state of the handle may change, and LGS\_BEHAVIOR how the LGS reacts to these changes. 

The history of a process's execution, called a \textit{trace}, is defined by a sequence of transitions each executed in response to a certain event from the \textit{alphabet} of the process. Processes such as the above two interact through events in the intersection of their alphabets, such as ``up'' and ``down''.

The concept of \textit{fluent} serves to express that a process may be in a certain state which it can only leave through specific transitions. For example, specifying 

{\footnotesize\begin{verbatim}
fluent HANDLE_IS_DOWN = <{down}, {up}>
fluent HANDLE_IS_UP = <{up}, {down}>
\end{verbatim}}

specifies the notion of the handle staying up and down in ``the handle has been pushed up and stays up'', from requirements R12bis and R22, and the dual property from R11bis and R21. Only the specified transitions can, in each case, set and unset the fluent.

Here are further fluents for LGS, some with more than one setting or resetting event:

{\footnotesize\begin{verbatim}
fluent DOOR_IS_CLOSING = <{start_closing}, {end_closing, open}>
fluent DOOR_IS_CLOSED = <{end_closing}, {open}>
fluent GEAR_IS_EXTENDING = <{start_extension}, {end_extension, start_retraction}>
fluent GEAR_IS_EXTENDED = <{end_extension}, {start_retraction}>
fluent GEAR_IS_RETRACTING = <{start_retraction}, {end_retraction, start_extension}>
fluent GEAR_IS_RETRACTED = <{end_retraction}, {start_extension}>
\end{verbatim}}

Some of the new events, such as start\_closing, do not immediately reflect an event expressed in the informal LGS specification, but are an artifact for expressing timing properties in the FPS framework.  

LTSA also supports assertions to express properties of the system. In the LGS example:

{\footnotesize\begin{verbatim}
assert R21 = [] ([] HANDLE_IS_DOWN -> [] ! GEAR_IS_RETRACTING)
assert R22 = [] ([] HANDLE_IS_UP -> [] ! GEAR_IS_EXTENDING)
assert R11bis = [] ([] HANDLE_IS_DOWN -> <> [] (GEAR_IS_EXTENDED && DOOR_IS_CLOSED))
assert R12bis = [] ([] HANDLE_IS_UP -> <> [] (GEAR_IS_RETRACTED && DOOR_IS_CLOSED))
\end{verbatim}}

with the following syntax for operators of boolean and temporal logic: ! is negation, \&\& is conjunction, -> is implication, [] is ``always'' and $\langle \rangle$ is ``eventually''.

The specification as given so far would not verify because of FPS's default assumption of equal priority of all applicable transitions,to ensure fairness. In R21, [] HANDLE\_IS\_DOWN cannot hold because an ``up'' event will eventually invalidate this fluent; similarly for R11bis. To resolve such situations, LTSA provides the  \textit{lower priority} operator, written $\rangle \rangle$. We can use it to resolve the conflict in favor of ``down'' by lowering the priority of ``up'', rewriting the definition of the system as
{\footnotesize\begin{verbatim}
||LGS = (LGS_BEHAVIOR || CONTROL_HANDLE) >> {up}.
\end{verbatim}}
and use the following auxiliary assertion to check the effect:
{\footnotesize\begin{verbatim}
assert EVENTUALLY_ALWAYS_DOWN = <> [] HANDLE_IS_DOWN
\end{verbatim}}
Then R21 and R11bis will verify but not R22 and R12bis, for which we would need instead to decrease the priority of ``down'' and assert EVENTUALLY\_ALWAYS\_UP. It is a characteristic of LTSA that in such cases one cannot verify both sets of assertions under the same conditions.

Appendix A contains the complete FSP model for the LGS example. The reader may input it ``as is'' into the LTSA analyzer.

Appendix \ref{sec:ltsa_appendix} contains the complete FSP model for the LGS example. An interested reader may put it into the LTSA analyzer as it is to do some experimenting.

Assessing LTSA according to the criteria of section \ref{sec:sec:criteria}:
\begin{itemize}
	\item \textit{Audience}: the FSP notation is mathematically formal and requires the corresponding qualification both from the specifiers and the readers.
	\item \textit{Abstraction}: the FSP abstraction is suitable both for specifying systems' implementations and for specifying requirements.
	\item \textit{Method}: the approach does not assume a particular requirements engineering method.
	\item \textit{Tool}: the LTSA tool supports model-checking FSP specifications, execution of specifications, and graphical simulation.
	\item \textit{Traceability}: the approach focuses on model-checking FSP specifications, so it offers no specific support for traceability.
	\item \textit{Coverage}: LTSA can specify and model-check liveness, safety, progress and fairness.
	\item \textit{Scope}: The approach can cover both system and environment aspects of requirements.
	\item \textit{Verification}: LTSA supports verification of liveness, safety, progress, and fairness properties.
	\item \textit{Semantics}: the FSP semantics is rigorously defined \cite{LTSA}. 
\end{itemize}

\subsubsection{FORM-L} \label{sec:sec:sec:form_l}

\begin{figure}[ht]
\includegraphics[scale=0.7]{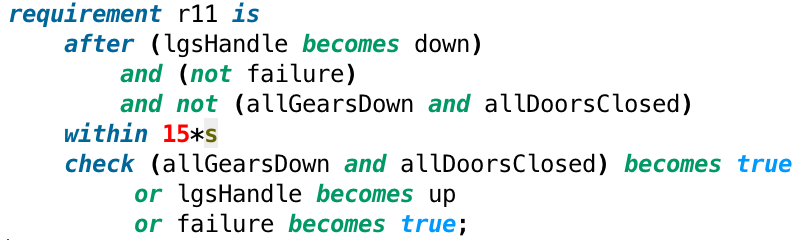}
\caption{Example of FORM-L requirement
%(taken from \cite{???})
}
\label{fig:forml}
\end{figure}

FORM-L \cite{nguyen_verification_2015} extends MODELICA \cite{Modelica}, an object-oriented notation for modeling the behavior of physical \textit{systems}. FORM-L results from the  MODRIO project (MOdel DRIven physical systems Operation), which improved MODELICA by adding the modeling of assumptions on the \textit{environment}. \myfig{forml} shows an example FORM-L specification.
%System vs environment

%Intended audience
%Level of abstraction
FORM-L addresses the early stage of system development, with a  level of detail sufficient to support some early validation through model-checking using the Stimulus tool \cite{Stimulus}.
%Associated method
%Traceability support
%Non-functional requirements
There is no particular method attached to FORM-L, also no support for traceability and non-functional requirements.
%Semantic definition
%Tool support
%Verifiability
Thanks to its formal semantics and to the support of a transformation (e.g., to , the FORM-L requirements can be verified by model-checking. 

Assessing FORM-L according to the criteria of section \ref{sec:sec:criteria}:
\begin{itemize}
	\item \textit{Audience}: the FORM-L notation is mathematically formal and requires the corresponding qualification both from specifiers and readers.
	\item \textit{Abstraction}: the FORM-L abstraction is suitable both for specifying systems' detailed design and for specifying goals and requirements.
	\item \textit{Method}: The approach does not assume any particular requirements engineering method but the main definition steps focus on: Goals, Requirements, Specification, Design.
	\item \textit{Tool}: Only tools internal to EDF, the organization that developed FORM-L, support code generation and simulation.
	\item \textit{Traceability} is among the goals but with no supporting mechanisms so far.
	\item \textit{Scope}: the approach can cover all aspects of requirements, including both system and environment aspects.
	\item \textit{Verification}: the approach  supports simulation.
	\item \textit{Semantics}: the FORM-L notation is in the process of being formalized by providing a formal semantics to a kernel set of FORM-L concepts.
\end{itemize}

\subsection{Other mathematical frameworks}
\label{sec:sec:sec:other_math}
%\TODO{Manuel first draft}
%\TODO{Sophie - polish text done - review done }
Requirements can use other mathematical theories other than graphs and automata, for example set theory, the ultimate basis for the two approaches discussed below, Event-B and Alloy. Other important mathematics-based approaches include VDM (Vienna Development Method), FSP/LTSA and Tabular Relations, both presented in the online version of this article.

\subsubsection{Event-B} \label{sec:sec:sec:event_b}

Event-B \cite{Abrial2010} is a formal method for system-level modeling and analysis. Modeling proceeds by specifying the system’s state in terms of sets and functions, and specifying state transformation in terms of  events. 
THe mathematical basis is set theory complemented by \textit{refinement}, a mechanism for turning a description of a system at a certain level of detail into a new one that remains consistent with it but includes more details. To ensure this consistency, a refinement must preserve the \textit{invariants} from the original description, while possibly adding new ones. The preservation of invariants must be proved mathematically, with the help of the supporting tools. 

\begin{figure}[ht!]
\includegraphics[scale=0.6]{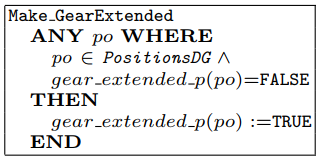}
\caption{Event to model gear extension (edited from \cite{Mammar2017})}
\label{fig:gextended}
\end{figure}

\begin{figure}[ht!]
\includegraphics[scale=0.6]{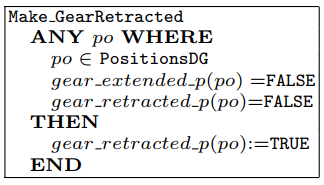}
\caption{Event to model gear retraction (edited from \cite{Mammar2017})}
\label{fig:gretracted}
\end{figure}

\begin{figure}[ht!]
\includegraphics[scale=0.65]{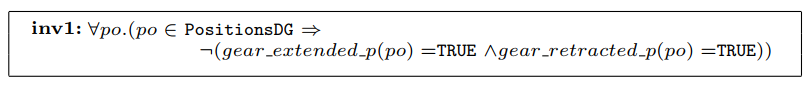}
\caption{Invariant: a gear cannot be extended and retracted at the same time (edited from \cite{Mammar2017})}
\label{fig:inv}
\end{figure}

Three fragments of an Event-B model for the LGS will illustrate the ideas.  \myfig{gextended} shows an event modeling the extension of the gear. The Boolean variable \textit{gear\_extended\_p} expresses whether the gear is extended or not. \myfig{gretracted} shows an event modeling the retraction of the gear. Another Boolean variable, \textit{gear\_retracted\_p}, models whether the gear is retracted or not. \myfig{inv} shows an invariant stating that the gear cannot be extended and retracted at the same time.

Assessing Event-B according to the criteria of section \ref{sec:sec:criteria}:
\begin{itemize}
	\item \textit{Audience}: The notation requires familiarity with classical set theory; while the concepts are elementary, they exclude stakeholders who feel uncomfortable with mathematics. 
	\item \textit{Abstraction}: Through successive refinement, the approach covers the full spectrum from very abstract and partial models to detailed final models ready for translation into an implementation. 
	\item \textit{Method}: beyond the notation, Event-B is a full-fledged method based on successive refinements proved correct (invariant-preserving). The method does not cover the entire process of requirements engineering (e.g. how to obtain requirements from stakeholders), only the refinement and proof process. 
	\item \textit{Tool}: tools support the refinement and proof process, particularly in the Rodin environment \cite{Rodin}.
	\item \textit{Traceability}: this is not a particular focus of the Event-B method, although it is possible to trace the identifiers used throughout the refinement process. 
	\item \textit{Coverage}: the approach makes it possible to model both a system and its environment. 
	\item \textit{Scope}: Event-B has no particular mechanism for modeling non-functional requirements. 
	\item \textit{Verification}: requirements expressed in Event-B are verifiable through the proof process which accompanies refinement: the description at every step of the refinement must be proved consistent with the description at the preceding (immediately higher) level. As a consequence, Event-B has been successfully used in several industrial projects requiring proofs of correctness, for example in the transportation and aerospace fields, and in business management \cite{Romanovsky2013}. In combination with other requirements approaches, such as Problem frames \cite{Jackson:2000}, it has also been used in the automotive industry \cite{Gmehlich2013}.
	\item \textit{Semantics}: 	%the approach is operational, and via the \textit{stepwise refinement} method more and more operational details are added to the model as long as the designer proceeds in following the methodological approach. The idea is to create a chain of models from abstract to implementation, from which code could be, in principle, generated.
the behavioral semantics of Event-B refinement has been described in \cite{Schneider2014}.
\end{itemize}

%===============VDM=================
\subsubsection*{4.4.A~Vienna Development Method} \label{sec:sec:sec:VDM}

The Vienna Development Method (VDM) \cite{BjornerJones78}, originally developed at the IBM Laboratory in Vienna in the 1970s, was one of the first formal methods in the history of software and system modeling. 
It includes the VDM Specification Language (VDM-SL) \cite{Overture} and its extended form (VDM++) \cite{Durr1992}. 
VDM-SL uses modules, while VDM++ applies object-oriented structuring with classes and multiple inheritance. 
Computing systems may be modeled in VDM-SL at a high level of abstraction, then transformed into progressively more detailed designs through a refinement process (\textit{reification}) similar to the Event-B ideas.

VDM-SL allows users to express different states of the systems and invariants that shall be met.
As an example, Figure~4.A~expresses requirements R21 and R22.

\begin{vdmsl}[label=lst:vdm-state,title={Figure 4.A. VDM-SL representation of the LGS states, with requirements R21 and R22}]
state LGS of
	gears: <retracted> | <extended> | <retracting> | <extending>
	doors: <opened> | <closed> | <opening> | <closing>
	handle: <down> | <up>
	inv mk_LGS(gears, doors, handle) ==
		(handle = <down> => gears <> <retracted>) and
		(handle = <up> => gears <> <extended>)
	init lgs == lgs = mk_LGS(<extended>,<closed>,<down>)
end
\end{vdmsl}

Another way to introduce requirements into a VDM specification is to translate the requirements into pre- and post-conditions of operations.
For example, the requirement R11bis (resp. R12bis) is a description of what happens when handle remains down (up).
Figure~4.B~specifies these operations; the \textbf{pre} clause expresses the condition under which the operation may be called, and the  \textbf{post} clause characterizes the state after execution.
\begin{vdmsl}[label=lst:vdm-operations,title={Figure 4.B VDM-SL operations specified by requirements R11bis and R12bis}]
operations
	extension_sequence() 
		ext wr gears
				 wr doors
				 rd handle
		pre handle = <down>
		post handle = <down> => (gears = <extended> and doors = <closed>);
		 
	retraction_sequence()
		ext wr gears
				 wr doors
				 rd handle
		pre handle = <up>
		post handle = <up> => (gears = <retracted> and doors = <closed>);
end LGS
\end{vdmsl}
The \textbf{rd} (read-only) and \textbf{wr} (read-write) clauses specify which parts of the LGS the operations may access and modify.

%by JMB: applying the common structure
Assessing VDM according to the criteria of section \ref{sec:sec:criteria}:
\begin{itemize}
	\item \textit{Audience}: VDM requires an understanding of mathematical foundations. 
	\item \textit{Abstraction}: VDM is a mathematical formalism, suitable for the early stages of the development process.
	\item \textit{Method}: VDM supports the description of data and functionality. Data are defined by means of types. Functionality is defined in terms of operations. 
	\item \textit{Tool}: VDM is mature and has an extensive tool support (e.g., Overture\cite{OvertureTool}). 
	\item \textit{Traceability}: VDM has no specific support for traceability. 
	\item \textit{Coverage}: The approach covers the expression of requirements on both the system and environment aspects. 
	\item \textit{Scope}: VDM covers the functional part of requirements. 
	\item \textit{Verification}: Supporting tools are available to verify VDM specifications.
	\item \textit{Semantics}: VDM has a formal semantics, published as an ISO (International Standards Organization) standard. 
\end{itemize}
%===============VDM=================

\subsubsection*{4.4.B~Process Algebra} \label{sec:sec:sec:process_algebra}

While Model-based formalisms such as Event-B and VDM are concerned with functional properties and sequential behavior, process algebras are concerned with interaction between concurrent processes. 
Among the original methods in this field, we can mention CSP \cite{Hoare:1978} and CCS \cite{Milner:1982}. 
Mobile process algebras (e.g. Milner’s $\pi$-calculus \cite{Milner:1982}) represent a further development by addressing channel mobility. 
The basic unit of all process algebras is a communicating concurrent process with its own input and output ports, which makes feasible modeling the interaction with the environment and expressing requirements beyond the system itself. 

%by JMB: applying the common structure
Assessing Process Algebra according to the criteria of section \ref{sec:sec:criteria}:
\begin{itemize}
	\item \textit{Audience}: the intended audience is that of specialist, mathematician and software engineer with deep understanding of foundations. 
	\item \textit{Abstraction}: Process Algebra can model a system at the level of abstraction of processes, regardless of the internal operational logic. 
	\item \textit{Method}: while Event-B is equipped with its own methodology base on refinement, Process Algebras are mainly a notation, although coming with notions of behavioral equivalence and techniques to formally prove processes semantical equivalence. 
	\item \textit{Tool}: the approach is not highly practical. 
Tool support is still limited: TyPiCal is a type-based static analyzer for the $\pi$-calculus \cite{Kobayashi}. 
TyPiCal is able to provide four different kinds of program analyses and transformations: lock-freedom analysis (certain communications or synchronizations will eventually succeed), deadlock-freedom analysis, useless-code elimination (it removes sub-processes that do not affect the observable behavior of the process), and information flow analysis. 
	\item \textit{Traceability}: requirements traceability is not supported. 
	\item \textit{Coverage}: among Process Algebra, CSP is possibly the one with the broadest set of concrete applications, typically in protocol analysis \cite{Lowe1996} or as a model for construction of parallel programming languages \cite{Hull1987}. 
Mobile Process Algebras such as the $\pi$-calculus have been instead mostly applied to the definition and formalization of service orchestration languages \cite{Lucchi05api-calculus,Mazzara:phd}.
	\item \textit{Scope}: non-functional requirements are not expressible. 
	\item \textit{Verification}: when both requirements and implementation are expressed in the form of a process, behavioral equivalence could be used to verify consistency. 
	\item \textit{Semantics}: all the algebras mentioned above come with their own formal semantics, often expressed via Structured Operational semantics (SOS). 

\end{itemize}

\subsubsection{Alloy} \label{sec:sec:sec:alloy}

Alloy \cite{Jackson:2006} is a declarative modeling language based on first-order logic for expressing the behavior of  software systems. It shares its roots with Event-B, since Alloy is a subset of the Z set-theory-based specification language \cite{Abrial:1980}, which also served as the starting point for Event-B.

%As with other math-based approach, users must possess some mathematical knowledge, here basic set theory; the syntax, however, is not standard mathematical notation but shows the influence of modeling languages such as UML \cite{Rumbaugh2004}). Alloy has formal syntax and semantics and brings to Z-style specification the automation offered by model checkers. Models can be automatically checked for correctness using the \textit{Alloy Analyzer}, a constraint solver that provides fully automatic simulation and checking based on a back-end SAT framework. Unlike Event-B, Alloy is a language with supporting does not come with an associated method like Event-B, and it is not intended for the description of non-functional properties, such as usability, performance, size, and reliability or intended for the description of timed behavior. Traceability of requirements is not natively supported.

Alloy has spurred a significant research community and a number of applications. \cite{Zave2017}, used Alloy to provide the  first specification of correct initialization and operations of the Chord ring-maintenance protocol \cite{Stoica2001}, using a formal model in Alloy and proving it correct.

In applying Alloy to the LGS example, we note that Alloy has no native mechanism for expressing properties such as ``handle remains in the DOWN position'' in R21. Alloy, however, can specify state transitions. Consider, however, this rewrite of R21:
\begin{description}
	\item [R21] If the landing gear command handle \textit{is} down, the gear is not retracting.
% 	\item [R22] If the landing gear command handle is up, the gear is not extending.
\end{description}

 with ``remains'' changed to ``is''. As an implication with a weaker antecedent, this new  R21 is stronger than the original. It can be expressed in Alloy:

\begin{lstlisting}[language=Alloy]
R21: check {
    all lgs, lgs': LGS | ((lgs.handle in Down) and (lgs'.handle in Down) and Main [lgs, lgs'])
        implies (lgs'.gear not in Retracting)
} for 5
\end{lstlisting}
where $Main$ is a predicate that yields ``true'' if and only if $lgs$ and $lgs'$ represent two consequent states of the LGS. The number 5 is the size of the search space, serving as bound for Alloy's use of bounded model checking for verification.

The absence of temporal operators in Alloy similarly suggests rewriting R11bis as:
\begin{description}
	\item [R11bis] If the handle remains down, three transitions of the LGS will suffice to have the gear extended and the door closed.
% 	\item [R12bis] If the handle remains up, three transitions of the LGS will suffice to have the gear retracted and the door closed.	
\end{description}
This form can be expressed in Alloy:
\begin{lstlisting}[language=Alloy]
R11bis: check {
    all lgs1, lgs2, lgs3, lgs4 : LGS |
        (((lgs1.handle in Down and lgs2.handle in Down and lgs3.handle in Down and lgs4.handle in Down) and
        (Main [lgs1, lgs2] and Main [lgs2, lgs3] and Main [lgs3, lgs4]))
            implies (lgs4.gear in Extended and lgs4.door in Closed))
} for 5
\end{lstlisting}
Three transitions involve four different states, which is why the assertion declares four variables of type ``LGS'' under the universal quantifier.

Assessing Alloy according to the criteria of section \ref{sec:sec:criteria}:
\begin{itemize}
	\item \textit{Audience}: the Alloy notation is mathematically formal and requires the corresponding qualification both from the specifiers and the readers.
	\item \textit{Abstraction}: the Alloy abstraction is suitable both for specifying systems' implementations and for specifying requirements.
	\item \textit{Method}: Alloy does not assume or promote a particular requirements engineering method.
	\item \textit{Tool}: the Alloy analyzer  supports model-checking of Alloy specifications, execution of specifications, and graphical simulation.
	\item \textit{Traceability}: Alloy focuses on model-checking specifications, so it offers no specific support for traceability.
	\item \textit{Coverage}: Alloy supports specifying and model-checking behavioral specifications, without consideration of non-functional properties.
	\item \textit{Scope}: Alloy can express both system and environment aspects.
	\item \textit{Verification}: Alloy models can be analyzed for consistency with counterexample-guided model-checking, and for  correctness of predicate-logic assertions with bounded model-checking.
	\item \textit{Semantics}: the Alloy semantics is rigorously defined in \cite{Jackson:2006}. 
\end{itemize}

We have specified and verified a complete LGS example in Alloy; it can be found in Appendix B.

Appendix \ref{sec:alloy_appendix} includes the complete Alloy example that we have specified and checked.

\subsubsection*{4.4.C.~Tabular relations} \label{sec:sec:sec:tabular_relations}

A general approach to requirements introduced by David Parnas \cite{parnas1992tabular} relies on specifying relations between system variables, represented in tabular form. The relations can be $n$-dimensional for any $n$. The underlying mathematical theory, relations over a collection of sets, is well known, and also serves as a basis for the relational model of databases.

Figures \hyperref[fig:r_11_12_bis]{4.C} and \hyperref[fig:r_21_22]{4.D} illustrate an application of the tabular approach to the LGS system. 
To capture requirements $R_{11}bis$ and $R_{12}bis$, since tabular representations do not support temporal operators, we introduce variables $gear\_status'$ and $door\_status'$, denoting the values of $gear\_status$ and $door\_status$ when the program finishes its execution.

The header of the table in Fig. \hyperref[fig:r_11_12_bis]{4.C} enumerates the possible conditions, where $NC (handle\_position)$ reflects the assumption that the position of the handle does not change \cite{parnas1993predicate}: $handle\_position = down \land NC (handle\_position)$, and $handle\_position = up \land NC (handle\_position)$. Requirements $R\_{21}$ and $R\_{22}$ talk about an immediate, rather than temporal, response of the system to the stimulus coming from the handle.

\begin{figure}[ht!]
\scriptsize
\begin{center}
\begin{tabular}{ | c || c | c | }
 \hline
  & $handle\_position = down \land NC(handle\_position)$ & $handle\_position = up \land NC(handle\_position)$ \\
 \hline\hline
 $gear\_status'$ & $extended$ & $retracted$ \\  
 \hline
 $door\_status'$ & $closed$ & $closed$ \\
 \hline
\end{tabular}
\end{center}
\captionsetup{labelformat=empty}
\caption{Fig.~4.C~Expressing requirements $R_{11}bis$ and $R_{12}bis$ with a vector function table \cite{parnas1992tabular}. The $NC(handle\_position)$ predicate states that the value of variable $handle\_position$ does not change \cite{parnas1993predicate}.}
\label{fig:r_11_12_bis}
\end{figure}

The resulting vector relation table in Fig. \hyperref[fig:r_21_22]{4.D} thus does not contain any additional variables.
An important property of a table is to be $proper$ --- that is, to characterize mutually disjoint situations.
The LGS knows two situations: when the pilots' handle in the cockpit stays (1) down and (2) up.
The LGS environment guarantees mutual disjointness of these situations.

\begin{figure}[ht!]
\scriptsize
	\begin{center}
		\begin{tabular}{ | c || c | c | }
			\hline
			& $handle\_position = down$ & $handle\_position = up$\\
			\hline\hline
			$gear\_status$ & $gear\_status \neq retracting$ & $gear\_status \neq extending$ \\  
			\hline
		\end{tabular}
	\end{center}
	\captionsetup{labelformat=empty}
    \caption{Fig.~4.D~Expressing requirements $R_{21}$ and $R_{22}$ with a vector relation table \cite{parnas1992tabular}. The table describes the relation connecting the handle's position and the gear's status.}
	\label{fig:r_21_22}	
\end{figure}

Assessing the tabular relations approach according to the criteria of section \ref{sec:sec:criteria}:
\begin{itemize}
	\item \textit{Audience}: the tabular notation is formal and requires mathematical qualification.
	\item \textit{Abstraction}: the notation operates at the level of program variables. Tables specify programs through specifying how the programs' executions affect the variables' values under mutually disjoint preconditions.
	\item \textit{Method}: the approach does not assume or promote a particular requirements engineering method.
	\item \textit{Tool}: while the approach, with its precise definition and simple tabular representation, would naturally lend itself to tool support, we found no record of such tools in the published literature.
	\item \textit{Traceability}: since tables directly use program variables, traceability to implementation would fit naturally, assuming appropriate tool support.
	\item \textit{Coverage}: the tabular notation focuses on specification of functional requirements. It defines ten classes of tables for this purpose \cite{parnas1992tabular}.
	\item \textit{Scope}: the approach can cover all aspects of requirements, including both system and environment aspects.
	\item \textit{Verification}: verification techniques that work with tabular relations include automated consistency checking 
	\cite{heitmeyer1996automated} and test oracle generation \cite{peters1998using}.
	\item \textit{Semantics}: the semantics of tabular relations is formally defined \cite{parnas1992tabular},
	\cite{parnas1993predicate}.
\end{itemize}

\subsection{Seamless, programming-language-based approaches} \label{sec:sec:pl_based}
% By Sasha
The following discussion of programming language-based approaches covers two recent developments: multirequirements method \cite{meyer_multirequirements_2013} and a refinement of it, Seamless Object-Oriented Requirements (SOOR) \cite{Naumchev2019SOOR, Naumchev2017}.

The underlying idea is the ``Single Model Principle'' described by Paige and Ostroff \cite{paige_single_2001} and going back to \cite{meyer1988object}. The principle holds that there is a fundamental unity in the software process, from requirements to design, implementation, validation and maintenance. Gaps between these tasks (``impedance mismatches'') are a threat to quality; to avoid them, it is desirable to use a seamless process with a consistent set of notations and tools throughout.
\subsubsection{Multirequirements} \label{sec:sec:sec:multirequirements}

%\item System vs environment
The multirequirements method treats requirements as compilable pieces of the program-to-be.
The pieces, being merged by special tools, form an initial program blueprint.
The method has limited consideration of environment-controlled properties: the resulting programs cannot assume anything about phenomena they cannot control.
It does makes it possible to constrain software-controlled environment phenomena: Design by Contract, on which the method relies, provides the necessary specification mechanisms.

%\item Intended audience
The method encourages the development of individual requirements incrementally over several layers including formal, graphical and natural-language.
The layers complement each other and are adapted to different shareholder needs, for example, respectively, precise descriptions advanced programmers and the quality-assurance team, communication and decision-making, and understanding by  non-software-technical user representatives.

%\item Associated method
Design by Contract techniques capture  behavioral aspects through preconditions, postconditions and class invariants, which (as a bonus) will remain part of the resulting program, helping a smooth interaction between requirements, design and implementation, and supporting traceability.

%\item Level of abstraction
The following principles of the multirequirements method reflect its abstraction level and the central role it assigns to the traceability goal:
\begin{description}
%The present section uses the exact lists' labeling from the original ``multirequirements'' article \cite{meyer_multirequirements_2013} to simplify searching.
\item[(3.3)] Model systems through object-oriented techniques: classes as the basic unit of decomposition, inheritance to capture abstraction variants, contracts to capture semantics.
\item[(3.4)] Use an object-oriented language (in the present discussion, Eiffel) to write the formal layer according to the principles of 3.3.
\item[(3.5)] Use the contract sublanguage of the programming language as the notation for the formal layer.
\item[(3.6)] As the goal is to describe models, not implementations, ignore the imperative parts of the programming language (such as assignment).
\item [(3.9)] Enforce and assess traceability between the layers and all products of the requirements process, and between requirements and other product artifacts, both down and up.
\end{description}

The method further refines the notion of traceability through the following definitions:
\begin{itemize}
\item \textit{Up-traceability}: every element of every artifact of a project follows from some element of the requirements.
\item \textit{Down-traceability}: for every requirement at least one artifact follows from it.
\end{itemize}

The Eiffel Information System (EIS) tool supports these complementary notions of traceability.

For verification, AutoProof \cite{tschannen_autoproof:_2015} supports static verification against requirements' contracts.

Applying the multirequirements method to requirements $R_{21}$ and $R_{22}$ yields the following formal  Eiffel representation:
\begin{lstlisting}
class LGS
invariant
  r21: (handle_position = down) implies (gear_status /= retracting)
  r22: (handle_position = up) implies (gear_status /= extending)
end
\end{lstlisting}
using assertion tags \texttt{r21:} and \texttt{r22:}, which is also useful for debugging.
For $R_{11}bis$ and $R_{12}bis$:
\begin{lstlisting}
class LGS
feature
  main
    do
    ensure
      r11_bis: (old handle_position = up and handle_position = up) implies (gear_status = retracted)
      r11_bis: (old handle_position = up and handle_position = up) implies (door_status = closed)   
      r12_bis: (old handle_position = down and handle_position = down) implies (gear_status = extended)
      r12_bis: (old handle_position = down and handle_position = down) implies (door_status = closed)         
    end  
end
\end{lstlisting}

\cite{naumchev_unifying_2016} contains a detailed application of the multirequirements method to a well-known example of a cyber-physical system.

\subsubsection{Seamless object-oriented requirements} \label{sec:sec:sec:seamless_requirements}
The seamless object-oriented requirements (SOOR) method \cite{naumchev2019exigences} extends the multirequirements approach by providing mechanisms to support requirements reuse and the specification of environment, temporal and timing properties. It contributes the following principles:
\begin{itemize}
    \item Use auxiliary routines with pre- and postconditions to specify and verify software behavior under different assumptions about the environment \cite{naumchev_complete_2016}.
    \item Inside these auxiliary routines, use loops equipped with loop invariants and variants to to specify and verify temporal and timing properties \cite{NAUMCHEV2019131}.
    \item Apply object-oriented genericity and abstraction techniques to make these routines universally reusable and to reduce their specification complexity \cite{DBLP:conf/tools/Naumchev19}.
\end{itemize}
The contracted auxiliary routines at the core of the method are called \emph{specification drivers} \cite{naumchev_complete_2016}.
They express formal properties of other features and classes; the technique is useful for adding contracts externally to uncontracted texts, or, if contracts are already present but partial, making the specification stronger.
Proving through Autoproof \cite{tschannen_autoproof:_2015} enables formal verification of correctness. 
In applications to testing, specification drivers can serve as parameterized unit tests (PUT's) \cite{Tillmann2005}.

As an illustration, the formal representation of the ``$R_{22}$'' requirementin SOOR will be:
\begin{lstlisting}
class
  R22
inherit
  ABSENCE_BEFORE [LGS, GEAR_EXTENDING, HANDLE_DOWN]
  LGS_REQUIREMENT
end
\end{lstlisting}

The \e{ABSENCE_BEFORE} class captures the pattern of the same name from a library of specification patterns for finite-state verification \cite{Dwyer1998}.
This pattern takes the following form in LTL:
\begin{equation}
    \Diamond R \Rightarrow (\neg P \cup R)
\label{eq:absence_before}
\end{equation}

The \e{R22} class applies this pattern to the \e{LGS} implementation class by inheriting from\\
\e{ABSENCE_BEFORE} with \e{LGS} for the implementation generic parameters, \e{GEAR_EXTENDING} for $P$ from Equation \ref{eq:absence_before} and \e{HANDLE_DOWN} for $R$ from the same equation.

The \e{ABSENCE_BEFORE [LGS, GEAR_EXTENDING, HANDLE_DOWN]} fragment reads as follows: \emph{in LGS, the GEAR is never EXTENDING until the HANDLE is pushed DOWN.}

The \e{ABSENCE_BEFORE} generic class already contains a specification driver capturing this property in a form suitable for either testing or program proving.
As a consequence, applying this pattern requires no manual writing of LTL formulae.
It amounts to inheriting from the \e{ABSENCE_BEFORE} class, providing system-specific properties as the generic parameters.
The \e{LGS_REQUIREMENT} class, from which \e{R22} inherits as well, captures LGS specific aspects.
Below is the specification driver from the \e{R22} class that captures the verifiable meaning of the requirement:
\begin{lstlisting}
frozen verify (system: LGS)
    -- (from ABSENCE_BEFORE)
  do
    from
      timer := time_boundary
    invariant
      p_does_not_hold_or_else_r_holds: not ({GEAR_EXTENDING}).default.holds (system) or else ({HANDLE_DOWN}).default.holds (system)
    variant
      timer
    until
      ({HANDLE_DOWN}).default.holds (system)
    loop
      iterate (system)
    end
  end
\end{lstlisting}
\cite{DBLP:conf/tools/Naumchev19} explains the technical details.
Reuse is an important part of the approach: template classes such as \e{ABSENCE_BEFORE}, free specifier from having to specify verification-related technicalities that apply to many specifications.
The \e{verify} routine is submittable to AutoProof, the Hoare-logic-based prover of Eiffel programs \cite{tschannen_autoproof:_2015}.
AutoProof will only accept the routine as correct if the \e{LGS} class has correct and strong enough contracts.

The SOOR representation of the ``$R_{12}bis$'' requirement takes the following form:
\begin{lstlisting}
class
    R12_BIS
inherit
    RESPONSE_GLOBAL [LGS, GEAR_EXTENDED_DOOR_CLOSED_OR_ELSE_HANDLE_DOWN, HANDLE_UP]
    LGS_REQUIREMENT
end
\end{lstlisting}
The \e{RESPONSE_GLOBAL} class captures the following LTL pattern of the same name \cite{Dwyer1998}:
\begin{equation}
    \Box (P \Rightarrow \Diamond S)   
\end{equation}
\e{RESPONSE_GLOBAL [LGS, ...]} expresses that \emph{in the LGS, when the HANDLE is pulled UP, the GEAR will eventually be seen EXTENDED and the DOOR will be seen CLOSED; OR ELSE, we conclude that the HANDLE is pushed back DOWN.}

Assessing both the multirequirements (Section \ref{sec:sec:sec:multirequirements}) and the SOOR (Section \ref{sec:sec:sec:seamless_requirements}) approaches according to the criteria of section \ref{sec:sec:criteria}:
\begin{itemize}
	\item \textit{Audience}: the three-layer representation yields complementary specifications, readable by different classes of stakeholders.
	\item \textit{Abstraction}: the use of programming-language notation makes it possible to cover the full spectrum from the most abstract requirements to the most concrete aspects of implementation.
	\item \textit{Method}: the approaches rest on a strong methodological basis, integrating the principles   object-oriented analysisthe followingLTL pattern of the same name \cite the EIS (Eiffel Information System) tool partially supports the process. It provides a mechanism for linking development objects o:rce pairwise traceability between requirements expressed in different notations by making the notion of a requirement multi-notational. EIS supports traceability between requirements and implementations.
	\item \textit{Coverage}: the approaches focus on functional requirements.
	\item \textit{Scope}: the approaches focus on system aspects. Environment properties could in principle be modeled in a similar way, but that aspect remains to be developed.
	\item \textit{Verification}: in the formal layer, requirements can be verified through tools such as AutoProof.
	\item \textit{Semantics}: the semantics of multirequirements comes from the semantics of contracts in  \cite{meyer2003framework}.
	\item \textit{Seamlessness}: along with traceability and verification, seamlessness is one of the highlights of these approach, which support carrying out the entire software development lifecycle using the same set of notations (the programming language) and tools (the IDE).
\end{itemize}

%----------------------------------------------------
\section{Results and discussion}\label{sec:Results}
%----------------------------------------------------

To draw conclusions from the present study, we first list limitations (\ref{ss:Limitations}). Then present a summarized table of results (\ref{ss:Results}). Finally we explore some of the conceptual questions raised by the notion of formal methods:

\begin{itemize}

\item Should the elicitation process start with an informal or semi-formal notation (\ref{ss:informalVs})?
\item Is a seamless approach better or worse than a mix of formal and semi-formal notations (\ref{ss:seamlessVs})?
\item What are the  merits of natural language and graphical notations for requirements (\ref{ss:Textual})?
\item What is the current state of tools support for requirements engineering (\ref{sec:tooling})?
\item What is the current state of education in formal approaches to requirements(\ref{sec:education})?
\end{itemize}

\subsection{Limitations}
\label{ss:Limitations}

While we have striven to make this review comprehensive, the following decisions may affect the generality of its results:
\begin{itemize}
\item The choice of a running example, the Landing Gear System. An alternative would have been to include a multitude of small examples, each chosen to make the corresponding approach shine, whereas the LGS may be more suitable to some than to others. The case for a single example is clear: to permit a significant comparison of the approaches. 
%MANUEL: I clarified that this is not always the case for every approach, but as a general statement it still stays!
\item The nature of  that example, a reactive system. An alternative would have been an enterprise-style system (accounting, Web content management, ...). The case for a reactive system is that such applications are among the hardest to build, so they are likely to test to their limits the advantages and deficiencies of the methods surveyed.
\item The LGS example, however, does not include concurrency, which prevents approaches such as Petri Nets or FSP/LTSA from showcasing some of their key properties.
\end {itemize}

\subsection{A summary of the results}
\label{ss:Results}

\mytab{results} presents the key conclusions in tabular form,  ordered by category (from  \mysec{approaches}), then alphabetically within each category.  For non-binary criteria, the table uses these notations:
\begin{itemize}
\item System vs Environment: either S (the approach can be only used to model the system) of  B (the approach can by used to model both system and environment).
\item Prerequisites: one of F (formal methods background, M (general mathematical knowledge), S (specific training required other than F and M),  N (no particular background expected)
\item Level of abstraction: one of L (low), H (high), B (both).
\end{itemize}

\begin{table}[hbt] 
\begin{adjustbox}{width=0.8\columnwidth,center}
\begin{tabular}{cl*{10}c}
        & \rot{System vs Environment} & \rot{Prerequisites} 
        & \rot{Level of abstraction} & \rot{Associated method} & \rot{Traceability support} 
        & \rot{Non-functional req. support} & \rot{Semantic definition}
        & \rot{Tool support} & \rot{Verifiability} \\ 
       %\cmidrule{2-12}

%--- Natural language
NL to OCL & S & S & H & \NOK & \NOK & \NOK & \OK & \OK & \OK  \\ 
NL to OWL & B & S & H & \OK & \OK & \OK & \OK & \NOK & \NOK  \\ 
NL to STD & B & S & H & \OK & \OK & \OK & \NOK & \NOK & \OK \\ 
Relax  & B & N & H & \NOK & \OK & \OK & \OK & \OK & \OK \\ 
Stimulus & B & N & H & \NOK & \NOK & \NOK & \OK & \OK & \OK  \\ 
Requirements Grammar & B & N & H & \NOK & \NOK & \OK & \NOK & \OK & \NOK  \\
%--- Semi-formal
Doors  & B & N & L &\NOK & \OK & \OK& \NOK & \OK & \NOK  \\ 
Reqtify  & B & N & L& \NOK & \OK & \OK & \NOK & \OK & \NOK \\ 
KAOS & B & S & B & \OK & \OK & \OK & \OK & \OK& \NOK \\ 
URN  & S & S & H & \OK & \OK & \OK & \NOK & \OK & \NOK \\ 
SysML  & S & S & H & \NOK & \OK & \OK & \NOK & \OK & \NOK  \\ 
URML  & B & S & H & \NOK & \OK & \OK & \OK & \OK & \NOK \\ 
%--- Graph and Automata
FORM-L  & B & S & H & \NOK & \NOK & \NOK & \OK & \OK & \OK \\ 
FSP/LTSA  & S & S & H & \OK & \NOK & \NOK & \OK & \OK & \OK  \\ 
Petri Nets  & S & S & H & \NOK & \NOK & \NOK & \OK & \OK & \OK \\ 
Problem Frames & B & S & H & \OK & \NOK & \NOK & \OK & \OK & \OK \\ 
Statecharts  & S & S & H & \OK & \NOK & \NOK & \OK & \OK & \OK \\ 
%--- Mathematical notation
Alloy & S & F & H & \NOK & \NOK & \NOK & \OK & \OK & \OK  \\ 
Event-B & B & F & B & \OK & \NOK & \NOK & \OK & \OK & \OK  \\ 
%Process Algebra & S & F & L & \NOK & \NOK & \NOK & \OK & \OK & \OK  \\ 
Tabular Relations & B & M & H & \OK & \NOK & \NOK & \OK & \NOK & \NOK  \\ 
VDM & B & F & H & \OK & \NOK & \NOK & \OK & \OK & \OK  \\ 
%--- Programming Language
Multirequirements  & B & S & B & \OK & \OK & \NOK & \OK & \OK & \NOK \\ 
       % \cmidrule[1pt]{2-12}
SOOR  & B & N & B & \OK & \OK & \NOK & \OK & \OK & \OK \\ 
       
\end{tabular}
\end{adjustbox}
\caption{Evaluation summary}
\label{tab:results}
\end{table}

\subsection{Informal versus semi-formal versus formal}
\label{ss:informalVs}

A requirements document should be both precise and understandable. These objectives can conflict with each other. Among the approaches surveyed,  formal methods favor precision at the possible risk of obscurity for non-experts; others, particularly natural-language-based and graphical, favor understandability, at the possible risk of renouncing precise semantics.

The issue of informality versus formality in the process of requirements engineering is not new. 
\cite{WongChengIn94} concludes that techniques providing a high degree of guidance and process description  are critical to achieve successful results. 
\cite{Lamsweerde:2000} concludes that higher-level abstractions for requirements specification and analysis are critical success factors. 

Formal methods have produced a number of industrial success stories, but their spread remains modest as assessed against the vast majority of projects using classical natural-language-based techniques. This survey may provide insight on how to extend that spread.

The usual argument against formal methods is that they are hard to understand. It has limits, however. Stakeholder comfort is a concern, but has to be matched against considerations of quality of the final system (``will the plane crash?''). Only a formal version can serve as a basis for a cohesive and unambiguous statement of client’s needs and, when necessary, for a binding legal contract as to what the system is supposed to do.

%however, the industrial use of formal techniques is still limited, and the process of adoption that would see them used by the average software engineer seems long. 
%Therefore, the question arises on whether the adoption of rigorous approaches is encountering so many obstacles to be deployed in the productive world, despite of the fact that so many influential researchers are strongly advocating in their favor. 
%We do not expect to be conclusive in our reflections here, or to give the ultimate answer to the question. However, we will collect our thoughts on the matter.

This counter-argument (in favor of formal methods) also has its limits. The rigor and precision of formal methods is not an excuse for ignoring the need to understand what stakeholders want. After all, even a system that has been formally ``proved correct'' has only been proved to \textit{satisfy a given specification}. However sophisticated the proof, if the specification does not reflect the stakeholders' desires, the system is in fact incorrect for all practical purposes. This observation is not just theoretical: numerous studies, most spectacularly by Lutz about NASA software \cite{DBLP:conf/re/Lutz93}, point to system failures resulting not from a technical error but from a bad understanding of user needs.

Any successful requirements method, formal or not, must provide good ways to understand and record stakeholders' intent. Unlike what a simplistic view might suggest, this process is not just a one-shot ``requirements elicitation'' phase but often, in practice, an iterative negotiation. Informal and graphical approaches have an advantage here since they are easy to explain to a broad range of stakeholders. To succeed on a large scale, formal methods and tools must provide similar mechanisms to interact with experts in the problem domain who are not experts in requirements. The discipline of requirements engineering traditionally recognizes (see textbooks such as \cite{wiegers2013software} and \cite{laplante2017requirements}) the need for ``requirements engineers'', also called ``business analysts'', who help translate needs as expressed by stakeholders, particularly ``domain experts'', into  bona fide requirements. To be successful for requirements elicitation, any formal method must develop its own cadre of such mediators, possessing both expertise in the method and an ability to relate to ordinary project stakeholders. Proponents of formal methods often complain about the reluctance of stakeholders to use mathematical reasoning. Complaining does not need anywhere and deflects from the formalists' own responsibility: never to start a formal-method-based requirements process without the right investment in requirements engineers who will translate back and forth between formal and informal views. 

%, we recognize the significance of understanding, capturing and recording the will and the original statements of the users. 
%We emphasize the negotiation with  stakeholders as the most important phase in the process. 
%Here, informality of requirements is important since it is not reasonable to expose the counterpart to a complex notation. 
%Requirements in natural language or in a graphical notation can be indeed easily discussed and validated by the average client.

%Secondly, we understand the importance of mapping these informal requirements into a formal representations with a higher level of rigor. 

The experience of the database community may provide guidance. In database design, the initial phase (before the switch to an implementation that often uses the relational model, another notation with a solid mathematical basis \cite{Codd:1970}), typically relies on a graphical semi-formal notation such as entity-relationships diagrams \cite{Chen:1976}, possessing a precise semantics but are intuitive enough for initial design and can be explained to non-expert stakeholders. This experience shows that, with a proper process in place, there is no reason to fear systematic rejection of formal or semi-formal approaches.

\subsection{Seamless versus conventional}
\label{ss:seamlessVs}
%\jmb{Currently proof-reading} => done

The dominant view in software engineering is that requirements and code are two fundamentally different products, to be handled through different methods, tools and languages. The drawback is the risk of divergence: software evolves, both on the requirements side and on the code size, and it is difficult to maintain consistency.

An alternative approach, discussed in section \ref{sec:sec:pl_based}, uses seamless development, relying on a single set of concepts and notations throughout; Eiffel in particular was designed as a language covering not only programming but also design as well as requirements. In such an approach there is a continuum from requirements to design to code, each step adding to the previously developed model, making it more concrete and closer to an actual program. One of the principal expected benefits is full traceability between requirements, design, code and other software artifacts. The ``multirequirements'' approach \cite{meyer_multirequirements_2013} extends the concept of seamlessness by using several complementary notations such as English, a formal notation or programming language, and a graphical notation, with the corresponding descriptions being kept in sync (the most formal of the versions serves as the reference).   
% conventional approach advocates disjointness of requirements and code, following the separation of concerns tradition. 
%The separation makes the two artifacts grow independently which often results in the two run out of sync. 
%The problem of tracing requirements to their implementations also gets more annoying as the development progresses. 
%Traceability links are often established manually, possibly with assistance of some configuration management tools. 
%Because requirements and code are not the only software process artifacts, -- unit tests represent another type, -- multiple interpretation of requirements becomes possible. 
%Different understanding of requirements by developers and testers represents a common situation in the software development routine. 
%Possible inability of developers and testers to agree on a requirement's meaning leads to involving the requirements analyst who has originally collected the requirement. 
%The project pays the price of the three parties' wasted time in the end.

The idea of using a programming language for requirements often triggers the reaction that programming languages are implementation-oriented and usually imperative, jeopardizing the necessary focus of requirements on ``what'' rather than ``how'' --- the Abstraction criterion of section \ref{sec:sec:criteria}. This concern, however, is not justified. Programming languages describe more than implementation. When applying them to requirements we may ignore their imperative aspects (although some of the authors' work does take advantage of imperative features \cite {naumchev2017contract, Naumchev2017, naumchev_unifying_2016, naumchev_complete_2016, DBLP:conf/tools/Naumchev19}). Language mechanisms developed over decades, particularly through object technology, have turned them into modeling tools for \textit{big} things. The notions of module/package, class, inheritance, information hiding, interface, genericity are examples of these contributions. While by definition the ``things'' being described are by default programs, these scaling-up techniques introduced by programming languages are applicable to the modularization of many other kinds of formal texts, including requirements.

These techniques coming from programming languages are in fact the only ones known to scale up to extremely large systems, such as a program of millions of lines of code. Ordinary mathematical notation is not designed for that purpose: mathematical statements typically extend over one or a few lines; to describe the relations between them, and the overall structure of a theory in an article or book, one has to resort to natural language. (This observation even applies to mostly formal mathematical texts such as Whitehead and Russell's \textit{Principia Mathematica}.) Notations for formal requirements, to be practical, need scaling-up capabilities; they can get them, ready for use, from programming languages with strong modular constructs. This is the vision behind Eiffel, with its full range of object-oriented modularization techniques, plus non-imperative specification techniques of Design by Contract, eminently applicable, beyond programs, to requirements of large systems.

Starting from the requirements-ready part of a programming language and retaining the same notation through the remaining software activities of design, implementation etc. presents the additional advantage of narrowing gaps (``\textit{impedance mismatches}'') between steps. The practical process of software development, whether ``waterfall'' or ``agile'' in principle, is inevitably back-and-forth in practice, with design and implementation forcing revisions of requirements. Too often issues or new ideas arising during implementation lead to changes that conceptually are \textit{requirements} change, but do not get reflected back into the requirements document or user stories because of the burden of converting back and forth between completely different frameworks and notations (typically a programming language for programs and natural language for requirements). If everything is in a single notation it becomes more realistic to keep everything in sync, with great advantages for traceability, debugging and maintenance. (This is the ``Single Model Principle'' discussed in \ref{sec:sec:pl_based}.)

%Without separation of concerns at all, managing the enormous complexity of the modern software projects would be impossible. 
%Nothing prevents, however, from refining the granularity. 
%The refinement has already happened on the activities' dimension: originally waterfall-like, software processes went agile. 
%In an agile process, development happens incrementally, with all the phases repeated in every iteration; the phases may also overlap in time within one iteration. 
%Why not scale the obvious success of agile to the artifacts' dimension? 
%That is, separate the representations at the level of individual units of knowledge, rather than at the level of whole artifacts. 
%This approach may align very smoothly with the modern tendency to make the process' granularity level finer. Building on the concept of seamless development, the `` multirequirements'' approach promotes using different, complementary formalisms, such as English, a formal notation or programming language, and a graphical notation, with the corresponding descriptions being kept in sync. 

%The seamless approach, as applied to requirements, pursues development of a specification notation that would bridge gaps between different software development artifacts.
%In this case, the following associated questions arise:
While seamless development runs contrary to the traditional emphasis on separation and concerns, it emphasizes the fundamental unity of software concepts throughout the lifecycle. In that view requirements are first-class citizens of the software world, on a par with other artifacts such as code, designs and tests, and susceptible to the same rules and techniques.

More work, in particular empirical, remains necessary to question and validate the seamless approach to formal and informal requirements:
\begin{itemize}
	\item Does seamlessness help make requirements useful for stakeholders with widely different backgrounds?
    \item What are the concrete traceability benefits?
	\item How much does seamless development reduce documentation overhead ?
	\item How much does it support requirements maintenance and reuse?
    \item Tools: in a seamless approach, are program development environments enough, or do requirements still call for specific tools?
    \item How can the processes of requirements and implementation reinforce each other?
\end{itemize}

\subsection{Textual versus graphical}
\label{ss:Textual}
% merged with:
%\subsection{User/Domain expert concerns}

In the discussion of how to make requirements understandable and expressible by various kinds of stakeholders, formal-versus-informal is not the only relevant criterion. Another opposition is textual versus graphical notations. The two distinctions are in fact orthogonal, as all four combinations exist:
\begin{itemize}
\item Informal specifications can be expressed in (textual) natural language, but they can also be graphical, in part or in full.
\item Formal specifications, often based on a textual mathematical notation, can also be expressed graphically. For example the presentations of Petri nets typically use a graphical form.  
\end{itemize}

While critics of formal requirements emphasize that stakeholders without a strong software or mathematical background can react negatively to formal texts, informal requirements are not necessarily the solution either: a long and verbose informal text can be just as off-putting. In contrast, graphical notations can make complex structures readily understandable. Among approaches covered in this survey, SysML, KAOS and i* are examples of methods (of various degrees of formality) that strongly and effectively rely on graphical notations. Another example, from the database community, was cited above: entity-relationship diagrams.

Graphics has clear advantages and limitations. A picture, it is said, is worth a thousand words. But it cannot carry the details of all these words. Graphical presentations are good at describing the overall scheme of a system --- what, with a revealing choice of words, is called ``the big picture''. For example, we can express graphically that an airplane guidance system has a component to control the trajectory and another to monitor it and raise alarms. Graphics is not, on the other hand, the best way to state the exact conditions (aircraft's altitude and angle) that will trigger an alarm.

As with formal vs informal approaches, textual and graphical notations are best viewed as complementary techniques, outside of any dogma, each to be used when and where it is the best way of specifying a given requirement element.

\subsection{Tool support} 
\label{sec:tooling}

Engineering the requirements of today's complex and ambitious systems cannot be a purely manual process.  Any realistic solution requires tool support. Modern requirements methods indeed come with tools, as cited in the previous sections. Graphical user interfaces, for example, are available in tools associated with methods ranging from the most informal to the fully formal, such as Reqtify \cite{Reqtify} for DOORS (\ref{sec:sec:sec:DoorsReqtify}), Objectiver \cite{Objectiver} for KAOS (\ref{sec:sec:sec:kaos}), Enterprise Architect \cite {EnterpriseArchitect} for SysML (\ref{sec:sec:sec:sysml}), Overture \cite{Overture} for VDM (\ref{sec:sec:sec:other_math}), AutoProof \cite{tschannen_autoproof:_2015}) for multirequirements in Eiffel \ref{sec:sec:sec:multirequirements}. Beyond user interfaces, however, what matters is how these tools help the requirements process. In particular:
\begin{itemize}
\item Which parts of requirements engineering do they facilitate?
\item How do they help achieve the core goal, requirements quality?
\end{itemize}

A 2011 survey of 94 requirements tools \cite{5929527} found that the emphasis was on modeling (42\% of the tools) and requirements management (39\%). This trend has continued. A commercial site listing the most widely used requirements tools\cite{Capterra} suggests that the principal functions of today's tools are requirements elicitation, change tracking and traceability. Reqtify, for example, provides functionalities to trace requirements and link requirements to artifacts of various kinds. 
The tools for KAOS \cite{Objectiver} and  i* focus on requirements elicitation. In addition to elicitation, tools for  
SysML \cite{EnterpriseArchitect} and URML help organize requirements into models and hierarchies, and the refinement process. All such tools, while a great aid to the requirements process and its integration with the rest or the development cycle, do not deal with formalization or deductive reasoning. 

Approaches based on seamlessness and the Single Model Principle \ref{sec:sec:pl_based} make it possible to rely on program proving tools to prove not only correctness properties of the future program (meaning its conformance to requirements) but also, at the requirements stage, consistency properties of the requirements themselves, independently of any forthcoming implementation. This is the approach taken in Eiffel with the AutoProof verification framework. Along with proofs, such an approach may benefit from modern tools for automated testing, such as AutoTest \cite{Meyer2009AutoTest}, Pex \cite{Tillmann2008} or AxiomMeister \cite{Tillmann2006Axiom}. Whether proof- or test-oriented, these tools need contracts as a basis for formal verification. 
%The notion of specification drivers, upon which the seamless requirements method relies, has a sibling in the world of testing:  the notion of parameterized unit tests (PUT's) \cite{Tillmann2005}. The main motivation behind the invention of PUT's was the need to abstract unit tests, without any direct connection to requirements. Although these two notions have a number of minor syntactical differences (in particular, PUT's lack frame conditions), they may reinforce each other's tool support. PUT's inhabit in the world of testing, which may explain the character of the two tools supporting them: Pex \cite{Tillmann2008} automatically generates unit tests based on PUT's preconditions using symbolic execution \cite{Tillmann2006}, and AxiomMeister \cite{Tillmann2006Axiom} discovers potential PUT's based on the source code.

%Moreover, a tool facilitating the expression of requirements through an attractive interface would respond to the needs of stakeholders. 

We note here a contribution of seamless development and the Single Product Principle (even if one does not follow it to its full extent) to the general discussion of tools for requirements: for maximum effectiveness \textit{tools supporting requirements should support more than requirements}. Consider the example of traceability. Some requirements tools have very good support for traceability between requirement elements. But traceability is also about tracing relations between requirements and other artifacts, particularly design, code and tests. (Traceability here involves detecting the consequences of a requirements change on all such possibly affected artifacts, and the other way around.) A tool focused just on requirements will not address this critical need. The future, we believe, lies in integrated tools that capture, along with requirements, all other products of software engineering. 

Open questions in the area of requirements tools include:
\begin{itemize}
\item Is there a general pattern for textual requirements, or does every domain area requires its own?
\item Can tools help measure the quality of requirements?
\item Should tools provide (in ``multirequirements'' style) different viewpoints tuned to each category of stakeholders?
\item Can functioning code be generated automatically from formal requirements? Should it?
\end{itemize}

\subsection{Education}
\label{sec:education}
% merge done by JMB
Interest in formal approaches to express requirements and, more generally, to design software has been progressively growing for the last few decades. 
Consequently, more and more educational institutions paired their research effort with a pedagogical effort in software engineering tracks \cite{Liu:2009}. 
This synergy, combined with the increasing interest of industry in formal approaches, grew a generation of students capable of developing formal thinking since the early stages of their professional career, and bringing this attitude to their job environments, being it a big corporation or their newly funded startup. 
The literature on pedagogical aspects of formal methods is vast \cite{Neville2009,tfm2009}, in particular on approaches to software engineering courses built on strong mathematical basis \cite{Gibson:1998}. 
Some resistance has been documented while deploying such approaches, and the issue of motivation has also been investigated \cite{Reed2004,ouhbi_requirements_2015}.

%It is far beyond the objectives of this work to be exhaustive to this regard. 
%We will only provide here an overview of general trends in formal method education, and describe examples of pedagogical use of the formalism and approaches described in this survey.

%\subsubsection*{General purpose}
The general purpose requirements formalisms are not often used to teach requirements engineering. 
If SysML is used, it is mainly in a larger scope of \gls{MBSE} education ( \cite{SysMlTuto}). 
Some other approaches like Kaos or i* are essentially used at requirements elicitation step (\cite{Nakatani:2008:REE:1566274.1566348}),
notably to students having no knowledge in formal methods (\cite{dalpiaz2015teaching}) or to complete formal method by domain specific ones (\cite{ishikawa_what_2009}). 
For example, there is a workshop series dedicated to i* teaching since 2015 (iStar@CAISE2015, iSTar@ER2017).

%\subsubsection*{NL-based}
The idea of natural-language-based approaches is, inter alia, to avoid the difficulty of teaching a new approach.
Indeed, by providing an interface between natural language and the formal representation of requirements, these approaches are based on the presupposed knowledge of natural-language requirements.
The teaching of expressing requirements with natural language is probably the first step of any requirements engineering course.
In some works \cite{hainey2011evaluation,zowghi_teaching_2003}, difficulties in teaching requirements elicitation are highligthed, and solutions are proposed to overcome them.
However, these difficulties are not linked to the use of natural language for requirements, but are about requirements qualities (such as completness, consistency, etc) and the emphasis is put on how to provide good requirements -- e.g., in \cite{wiegers2013software}, the authors propose ``good pratices'' for requirements.

%\subsubsection*{Graph and automata}

%\mm{JM?}
Graph and automata have always been a popular representation for students \cite{Grinder:2002:AAC:563517.563364,doi:10.1080/00207390903372429}. 
Despite their mathematical foundations, they are graphical, easy to understand, and not difficult to produce.
This apparent user-friendliness often leads to misunderstanding and errors because of the lacks of clear execution semantics, as reported by famous articles \cite{vonderBeeck1994,Cuccuru:2007:EUE:2394101.2394127,10.1109/MC.2006.65}.  
In that matters, the recent progress in executable semantics and in tools animation (see previous \mysec{tooling}) is a very important progress with regards to education.

%\subsubsection*{Other Mathematical}

In \cite{CatanoR09} experiences are reported in teaching formal methods, in particular JML \cite{Leavens98jml} and B \cite{Abrial:1996}. 
Design and delivery of courses aiming at developing skills of model construction and analysis by use of notations such as VDM-SL and VDM++ are presented in \cite{Larsen2009}. 
The motivation problem has been here improved by using examples from industrial projects, and by using an industrial-strength tool set. 
Concurrency theory and FSP/LTSA have also been used in teaching and documented \cite{Aceto2009}. 
Among all formalisms for concurrency, CSP has been the one with more widespread applications, both industrial and educational \cite{Roscoe:1997}.
More interactive ways of introducing formal specifications have been experimented, for example in \cite{Tarkan2009}, where an on-line tutorial has been designed to help students in the transition from Z to Alloy, considered the latter more practical to use due to the existence of the Alloy Analyzer. 

%\subsubsection*{Programming language-based}
%\mm{Sasha?}
Seamless, programming-language-based approaches to requirements are not yet widespread, so there is little empirical data available on their suitability for teaching. But a few observations are possible. Since Design by Contract (DbC) is the basis, the transition to a similar approach for requirements will be easy to explain to an audience which has been introduced to programming using DbC; this is the case with both approaches mentioned in Section \ref{sec:sec:pl_based}: with the multirequirements method, requirements take the form of contracted excerpts from the final program; with seamless requirements, they take the form of contracted routines expressed in terms of the final program. 
%Speaking about the seamless requirements method, it may also take advantage of existing integrated development environments (IDE's) since requirements documents become compilable classes inside of the software projects.
%
% These speculations on programming language-based approaches take us to a natural conclusion that the best way to teach them would be in the form of extensions to existing programming courses that rely on \gls{DBC}. 
% In this regard such approaches bring seamlessness not only to software engineering itself, but also to the way it is being taught.

On education, a number of questions remain open:
\begin{itemize}
\item Are extensions to existing courses relying on \gls{DBC} a suitable approach to teach both programming and formal methods with an emphasis on quality?
\item Should software engineering courses emphasize seamlessness?
\item Why is there so little emphasis on requirements in regular Software Engineering curricula?
\item How much should formal notations appear in introductory courses?
\end{itemize}

\subsection{Conclusion: what role for formal approaches to requirements?}

What degree of formality is appropriate in stating requirements for software systems? To shed light on that question, this survey has analyzed a wide range of techniques for expressing software requirements, with a degree of formalism ranging over a broad scale: from completely informal (natural language), through partially formal (semi-formal, programming-language-based), to completely mathematical (automata theory, other mathematical bases). 

The question of formality has caused and continue to cause heated debates, almost as old as the very recognition of requirements engineering as a  significant component of software engineering. In those discussions, the basic arguments for and against have not changed much over  decades: inevitably, proponents of formal methods will point to the imprecision of natural language; just as inevitably, opponents will argue that formal texts are incomprehensible to many stakeholders. There is truth is such statements on both sides, but they cannot end the discussion. The detailed analysis and examples of this article should help reach better informed decisions.

As an example of the limits of classic but simplistic views, consider the ``many stakeholders do not understand formal notations'' argument. In reality, no one can require that all stakeholders understand all details of requirements. There is no such rule in other engineering endeavors; the marketing manager for a car company, perhaps the primary stakeholder since what counts is how cars will sell, cannot understand all the engineering diagrams and technical decisions. Even if we limit our focus to software, many aspects of any sophisticated software system will remain impenetrable to \textit{some} stakeholders: if the system's scope extends across many technical areas, as in the case of a banking system that touches on accounting, investment management, currency handling, international transfers etc., no single stakeholder is an expert in all these disciplines.

One suspects that often that the formal-is-hard argument is really formal-is-hard-for-\textit{my}-\textit{developers}. People who, for example, promote semi-formal Design by Contract techniques regularly hear such comments: this is too hard for our people, they would need retraining, or maybe they just do not have the right mathematical education. Such objections are worth considering, but raise questions: why do these concerns matter more than others such as verifiability of the requirements? Comparing again with other areas of engineering, a building contractor is unlikely to use as an excuse, if the circuits short, that he could not require his electricians to learn Ohm's law.

Beyond simplistic arguments, we need a balanced view assessing formality against relevant \textit{criteria} of quality. This is the focus of the present article: each of the reviewed approaches has been evaluated according to a set of criteria introduced in \ref{sec:sec:criteria}. These criteria, while not the only possible ones, are intended to cover what is most important to the stakeholders of a system. 

In light of that review, several observations serve as a counterweight to ``formalism-is-hard'':
\begin{itemize}
\item Stakeholders who do not understand a formal description of a system still need to understand many of its aspects (as the car marketer must understand what is new in the latest model). The notion of \textit{view} is useful here. Requirements for a system may have to rely on several views adapted to the needs of different stakeholders; examples include natural language, graphical, tabular views and of course a formal view, the appropriate one for stakeholders who need precision and must consequently be ready to deal with mathematical concepts. (Mathematics is not a torture imposed on innocent stakeholders; it is the only language with full precision and, as a consequence, the language of science.) From this perspective, a formal expression of the requirements does not compete with other variants, but complements and supports them.
\item If requirements use multiple views, the question arises of how to guarantee that they are consistent; only a formally defined view has the rigor and precision needed to be usable as the basis to derive others. The multirequirements method \cite{meyer_multirequirements_2013} develops this idea further, proposing to write requirements in a combination of natural-language, graphical and formal notations, the formal one expressed in Eiffel and serving as the reference in case of ambiguity. 
\item For most practical uses, the level of mathematics actually required to understand formal descriptions, and even in many cases to write them is not particularly high. Many software engineers and other professionals have gone through science curricula in which they had to master challenging mathematical techniques, such as control theory and statistics. For most formal methods the underlying mathematics consists of basic set theory and basic logic in the form of propositional and predicate calculus. (Specifications of real-time systems may also use temporal logic, but it is a simple extension to logic and not hard to learn.) The difficulty is often apparent rather than real; a matter of attitude.
\item Anyone working in software is used to highly formalized (although usually not mathematical) notations: programming languages, which leave no room for imprecision.
\item Executable semantics for general techniques such as UML or SysML have enjoyed widespread use, showing that when benefits are clear users willlearn highly technical approaches.
\end{itemize}

The last comment suggests a way to progress in the formal-versus-informal debate. Ideological discussions should yield to pragmatic considerations. Techniques will gain acceptance if they produce tangible benefits commensurate with the effort they require. Two crucial conditions are:
\begin{itemize}
\item Tools: even the most impressive method and elegant notation will not catch on without automated support. Good tools free programmers from mundane tasks, flag errors and inconsistencies, and scale up to large systems.
\item Education: software engineering education often causes disconnects where it should emphasize synergy. Disconnect between requirements and subsequent tasks, particularly implementation. Disconnect between formal methods, often taught as a special advanced topic for theory-inclined students, and the practice of software engineering (section \ref{ss:seamlessVs} discussed the arguments for a more \textit{seamless} approach, which threads these tasks together).

%The educational process often tends to create an artificial separation between the definition of requirements and their implementation. For example, often requirements are given by teacher and implemented by students. This creates an unnecessary wall in terms of perceptions of what the responsibilities of a good software engineer should be, and prevents students to practice the discipline. We advocate the importance of a seamlessness approach also in the educational process, i.e. students should be trained to define requirements and to realize how important it is to write them properly.
\end{itemize}

Even these regrets about disconnects between courses rely on an optimistic assumption: that students take courses on requirements and courses in formal methods. It is in fact possible today to complete a computer science/informatics/software engineering curriculum without having had courses on both of these topics --- or, in some cases, on either of them. Such curricula should be corrected: every software engineer needs to know about requirements engineering, the discipline of making sure that the implementation of systems meets the needs of their stakeholders and the constraints of the environment; every software engineer should know how to apply formal techniques when precision and guaranteed correctness are required; and every software engineer should know when and how requirements can benefit from formal methods. 

Beyond their application to education, these observations describe the relationship between formal methods and requirements in software engineering. Formal methods are sometimes considered theoretical while requirements engineering is essential to the practice of software construction. For that practice, formal methods complement other requirements techniques rather than attempting to replace them. They can and should be a powerful help available to every requirements engineer or business analyst.

We hope that the present survey has demonstrated this potential contribution of formal methods to requirements. We also hope that it will contribute to expanding their role for the greater benefit of future software systems and the people who depend on them.

\subsection*{Acknowledgments}
We are grateful to the authors of the surveyed approaches who took time to check our rendering of their approaches and the treatment of the LGS examples. (All responsibility for these descriptions remains ours.) They include in particular Jeff Kramer, who went so far as to develop his own LTS specification of the LGS example from which our specification benefited considerably, and for his insightful comments, and Thuy Nguyen for his formalization of the LGS in Form-L.

%----------------------------------------------------
\bibliographystyle{ACM-Reference-Format}
\bibliography{biblio,surveys}

\appendix

\section{LGS in FSP/LTSA} \label{sec:ltsa_appendix}
\scriptsize{
\begin{verbatim}
// Reaction of the LGS to the different states of the handle.
LGS_BEHAVIOR = (do_nothing -> LGS_BEHAVIOR | up -> OPEN_FOR_RETRACTION | down -> OPEN_FOR_EXTENSION),
OPEN_FOR_RETRACTION = (open -> START_RETRACTION),
START_RETRACTION = (start_retraction -> END_RETRACTION),
END_RETRACTION = (end_retraction -> START_CLOSING_RETRACTED | down -> START_EXTENSION),
START_CLOSING_RETRACTED = (start_closing -> END_CLOSING_RETRACTED),
END_CLOSING_RETRACTED = (end_closing -> LGS_BEHAVIOR | down -> OPEN_FOR_EXTENSION),
OPEN_FOR_EXTENSION = (open -> START_EXTENSION),
START_EXTENSION = (start_extension -> END_EXTENSION),
END_EXTENSION = (end_extension -> START_CLOSING_EXTENDED | up -> START_RETRACTION),
START_CLOSING_EXTENDED = (start_closing -> END_CLOSING_EXTENDED),
END_CLOSING_EXTENDED = (end_closing -> LGS_BEHAVIOR | up -> OPEN_FOR_RETRACTION).
// Fluents that model possible states of the LGS:
fluent HANDLE_IS_DOWN = <{down}, {up}>
fluent HANDLE_IS_UP = <{up}, {down}>
fluent DOOR_IS_CLOSING = <{start_closing}, {end_closing, open}>
fluent DOOR_IS_CLOSED = <{end_closing}, {open}>
fluent GEAR_IS_EXTENDING = <{start_extension}, {end_extension, start_retraction}>
fluent GEAR_IS_EXTENDED = <{end_extension}, {start_retraction}>
fluent GEAR_IS_RETRACTING = <{start_retraction}, {end_retraction, start_extension}>
fluent GEAR_IS_RETRACTED = <{end_retraction}, {start_extension}>
// Specifying the LGS control handle.
CONTROL_HANDLE = (down -> INITIALLY_DOWN | up -> INITIALLY_UP),
INITIALLY_DOWN = (up -> down -> INITIALLY_DOWN),
INITIALLY_UP = (down -> up -> INITIALLY_UP).
// Modelling different modes of operation.
||LGS = (LGS_BEHAVIOR || CONTROL_HANDLE) >>
{up}. // Uncomment to model LGS with the handle pushed down.
//	{down}. // Uncomment to model LGS with the handle pulled up.
assert EVENTUALLY_ALWAYS_DOWN = <> [] HANDLE_IS_DOWN
assert R11bis = [] ([] HANDLE_IS_DOWN -> <> [] (GEAR_IS_EXTENDED && DOOR_IS_CLOSED))
assert R21 = [] ([] HANDLE_IS_DOWN -> [] ! GEAR_IS_RETRACTING)
assert EVENTUALLY_ALWAYS_UP = <> [] HANDLE_IS_UP
assert R12bis = [] ([] HANDLE_IS_UP -> <> [] (GEAR_IS_RETRACTED && DOOR_IS_CLOSED))
assert R22 = [] ([] HANDLE_IS_UP -> [] ! GEAR_IS_EXTENDING)

\end{verbatim}
}

\section{LGS in Alloy} \label{sec:alloy_appendix}

\begin{lstlisting}[language=alloy]
abstract sig Handle {}
sig Down, Up extends Handle {}{Down + Up = Handle}
abstract sig Door {}
sig Closed, Opening, Open, Closing extends Door {}{
	Closed + Opening + Open + Closing = Door
}
abstract sig Gear {}
sig Extended, Retracting, Retracted, Extending extends Gear {}{
	Extended + Retracting + Retracted + Extending = Gear
}
sig LGS {
	handle: Handle,
	door: Door,
	gear: Gear
} {
	(gear in Retracting || gear in Extending) implies (door in Open)
	(handle in Up) implies (gear not in Extending)
}
pred closeDoor [lgs, lgs' : LGS] {
	(lgs.door in Open) implies (lgs'.door in Closing)
	(lgs.door in Closing) implies (lgs'.door in Closed)
	(lgs.door in Opening) implies (lgs'.door in Closing)
	(lgs.door in Closed) implies (lgs'.door in Closed)
	(lgs.gear = lgs'.gear)
}
pred openDoor [lgs, lgs' : LGS] {
	(lgs.door in Open) implies (lgs'.door in Open)
	(lgs.door in Closing) implies (lgs'.door in Opening)
	(lgs.door in Opening) implies (lgs'.door in Open)
	(lgs.door in Closed) implies (lgs'.door in Opening)
	(lgs.gear = lgs'.gear)
}
pred retractGear [lgs, lgs' : LGS] {
	(lgs.gear in Extended) implies (lgs'.gear in Retracting)
	(lgs.gear in Retracting) implies (lgs'.gear in Retracted)
	(lgs.gear in Extending) implies (lgs'.gear in Retracting)
	(lgs.gear in Retracted) implies (lgs'.gear in Retracted)
	(lgs.door = lgs'.door)
}
pred extendGear [lgs, lgs' : LGS] {
	(lgs.gear in Extended) implies (lgs'.gear in Extended)
	(lgs.gear in Retracting) implies (lgs'.gear in Extending)
	(lgs.gear in Extending) implies (lgs'.gear in Extended)
	(lgs.gear in Retracted) implies (lgs'.gear in Extending)
	(lgs.door = lgs'.door)
}
pred retractionSequence [lgs, lgs' : LGS] {
	(lgs.gear not in Retracted) implies openDoor [lgs, lgs']
	((lgs.gear not in Retracted) && (lgs.door in Open)) implies retractGear [lgs, lgs']
	(lgs.gear in Retracted) implies closeDoor [lgs, lgs']
}
pred outgoingSequence [lgs, lgs' : LGS] {
	(lgs.gear not in Extended) implies openDoor [lgs, lgs']
	((lgs.gear not in Extended) && (lgs.door in Open)) implies extendGear [lgs, lgs']
	(lgs.gear in Extended) implies closeDoor [lgs, lgs']
}
pred Main [lgs, lgs' : LGS] {
	(lgs.handle in Up) implies retractionSequence [lgs, lgs']
	(lgs.handle in Down) implies outgoingSequence [lgs, lgs']
}
run Main for exactly 2 LGS, 2 Handle, 4 Door, 4 Gear
R21: check {
	all lgs, lgs': LGS | ((lgs.handle in Down) and (lgs'.handle in Down) and Main [lgs, lgs']) implies (lgs'.gear not in Retracting)
} for 5
R22: check {
	all lgs, lgs': LGS | ((lgs.handle in Up) and (lgs'.handle in Up) and Main [lgs, lgs']) implies (lgs'.gear not in Extending)
} for 5
R11bis: check {
	all lgs1, lgs2, lgs3, lgs4 : LGS |
		(((lgs1.handle in Down and lgs2.handle in Down and lgs3.handle in Down and lgs4.handle in Down) and
		(Main [lgs1, lgs2] and Main [lgs2, lgs3] and Main [lgs3, lgs4]))
			implies (lgs4.gear in Extended and lgs4.door in Closed))
} for 5
R12bis: check {
	all lgs1, lgs2, lgs3, lgs4 : LGS |
		(((lgs1.handle in Up and lgs2.handle in Up and lgs3.handle in Up and lgs4.handle in Up) and
		(Main [lgs1, lgs2] and Main [lgs2, lgs3] and Main [lgs3, lgs4]))
			implies (lgs4.gear in Retracted and lgs4.door in Closed))
} for 5
\end{lstlisting}

\end{document}